\documentclass[final,5p,times,twocolumn]{elsarticle}

\usepackage{amsmath,amssymb,amsthm,bm,mathtools}
\usepackage{booktabs,multirow,graphicx}
\usepackage{algorithm,algpseudocode}
\usepackage[hidelinks]{hyperref}
\graphicspath{{../}}

\newtheorem{theorem}{Theorem}
\newtheorem{proposition}{Proposition}

\theoremstyle{definition}

\DeclareMathOperator{\tr}{tr}
\DeclareMathOperator{\med}{med}
\DeclareMathOperator*{\argmax}{arg\,max}
\DeclareMathOperator*{\argmin}{arg\,min}
\newcommand{\R}{\mathbb{R}}
\newcommand{\E}{\mathbb{E}}
\newcommand{\Prob}{\mathbb{P}}
\newcommand{\lse}{\operatorname{lse}}
\newcommand{\T}{\top}
\newcommand{\bS}{\bm{\Sigma}}
\newcommand{\hbS}{\widehat{\bm{\Sigma}}}
\newcommand{\ckpt}[1]{{\scriptsize\texttt{#1}}}

\journal{Elsevier}

\begin{document}

\begin{frontmatter}

\title{Knowledge Fermentation: Learning from Shared Variation in Few-Shot Classification}

\author[1]{Ali Mohammadi Ruzbahani}
\ead{ali.mohammadiruzbaha@ucalgary.ca}
\author[2]{Abbas Yazdinejad}
\ead{Abbas.Yazdinejad@uregina.ca}
\author[1]{Hadis Karimipour}
\ead{hadis.karimipour@ucalgary.ca}
\address[1]{Smart Cyber-Physical Systems (SCPS) Lab, University of Calgary, Calgary, Alberta, Canada}
\address[2]{Decentralized Cybersecurity and Artificial Intelligence (DCAILab), University of Regina, Regina, Saskatchewan, Canada}

\begin{abstract}
A classifier built from a handful of labeled examples knows where each class sits in feature space but not how it varies, which is why prototype rules and linear probes overfit in the low-data regime. Knowledge Fermentation (KF) recovers this second-order knowledge by letting classes exchange what they know about variation. Residuals around class prototypes act as carriers; the variance of each direction is measured on classes that took no part in estimating it, and directions above the level of the surrounding medium define a structured noise that acts on a linear classifier as a closed-form change of geometry. A convex objective trains the classifier at a noise strength calibrated on the distances between prototypes, and unlabeled examples, when available, join as carriers in repeated rounds. We prove that the held-out variances are unbiased for the directions on which they are measured, that the fermented map never expands a direction, and that damping caused by wrong assignments limits itself. On 14 datasets from vision and language and 8 frozen encoders, with all constants fixed, KF is on par with the best of six tuned inductive baselines. With unlabeled carriers it raises one-shot accuracy from 52.4\% to 65.8\% on average and attains the highest average accuracy among nine transductive methods, significantly so on vision encoders, while information maximization stays ahead on sentence encoders. Controlled studies attribute the gain to the task-level geometry and show that it grows with the within-class variation that the classes of a task share.
\end{abstract}


\end{frontmatter}

\section{Introduction}
\label{sec:intro}

A classifier that must be built from one or a few labeled examples per class fails for a specific reason: the examples reveal where a class is, but not how it varies. A frozen encoder maps every example to a point that mixes class identity with nuisance factors such as viewpoint, background, style or phrasing. With so few points per class, neither a prototype nor a linear probe can separate the two, because the directions along which features scatter inside a class cannot be observed from one class at a time. Estimates of that scatter are degenerate with a single example and dominated by sampling noise with a few, and the classifier ends up fitting the accidents of the support set. A survey of the field describes this as the unreliability of the empirical risk minimizer when supervision is scarce \cite{wang2021generalizing}; in feature space it has a concrete form, which is ignorance of the within-class covariance.

\begin{figure*}[t]
\centering
\includegraphics[width=\textwidth]{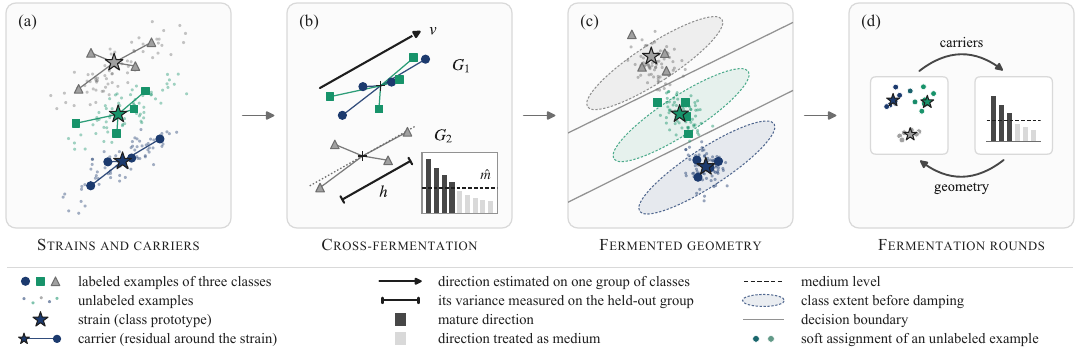}
\caption{Overview of Knowledge Fermentation. (a) Every class keeps a strain, the prototype of its labeled examples, and contributes carriers, the residuals around it. (b) Carriers are pooled within groups of classes: directions of variation $v$ are estimated on one group, their variance $h$ is measured on the held-out group, and directions above the medium level $\hat{m}$ are mature. (c) The fermented map damps the mature directions to the medium level, and a linear classifier is trained in the resulting geometry. (d) With unlabeled examples, soft assignments supply further carriers and the geometry is matured again in every round.}
\label{fig:overview}
\end{figure*}

Knowledge Fermentation (KF) solves this problem because it supplies exactly the missing quantity, and it does so without a single additional label. The method treats within-class variation as knowledge that classes hold in common and can exchange. Residuals of all classes around their prototypes are pooled as carriers of variation. The variance of every direction of variation is measured on classes that played no part in estimating it, and a direction is accepted as mature only if that measurement stands out from the surrounding medium, which filters out the directions that merely reflect which examples happened to be drawn. Mature directions are then turned into structured noise on the feature space. For a linear classifier the effect of this noise has a closed form: it is a change of geometry in which every mature direction is damped down to the level of the medium. A convex objective finally trains the classifier in this geometry under isotropic noise, at a strength calibrated so that a noisy prototype crosses to any competing class with a prescribed, small probability. If unlabeled examples from the same task are present, they become additional carriers, and the same exchange is repeated in rounds. Figure~\ref{fig:overview} summarizes the pipeline.

The name is borrowed from the knowledge fermenting model of organizational learning \cite{wang2008kfquality}, in which initial knowledge, the knowledge sourdough, develops through the nourishment of a knowledge matrix and the catalysis of a knowledge enzyme that promotes the integration of knowledge. We read the learning problem in these terms. Prototypes are the sourdough, the strains that fix what each class is. Residuals form the matrix in which knowledge about variation is stored. Validation of one group of classes against another plays the role of the enzyme, since it is the step that integrates what separate classes know. Unlabeled examples are further material, and each round repeats the fermentation.

Feature-space augmentation has been studied before, and the difference to prior work lies in where the statistics come from and how they are used. Hallucination and distribution calibration transfer variation from base classes with abundant data \cite{hariharan2017lowshot,yang2022bridging}, which a task with no base classes does not have. Implicit semantic augmentation estimates one covariance per class from that class alone and optimizes an upper bound of the augmented loss \cite{wang2021isda}, which is well suited to large training sets and loses its footing when a class has two examples. KF needs neither base classes nor many examples per class. It also replaces a hand-set augmentation strength by a rule derived from a guarantee, and sampling by a closed-form change of geometry.

The contributions of this paper are the following.
\begin{itemize}
\item A formulation of low-data classification as fermentation of second-order knowledge: carriers pooled across classes, a spectrum measured on held-out classes, and one-sided damping toward a medium level that the data determine. The held-out variances are proved unbiased for the directions on which they are measured, in contrast to in-sample eigenvalues, and the cost of transferring them to the directions of the full sample is bounded.
\item A fermented geometry with exact properties. For a linear classifier, structured noise on the encoder features is isotropic noise after the fermented map; the map whitens the matured covariance, never expands a direction, and is admitted at a level selected by an integrity gate that tests identification on held-out classes.
\item An integrity-calibrated convex objective. The noise strength follows in closed form from the probability that a fermented prototype crosses to a competing class, and the objective is shown to interpolate between empirical risk minimization and a prototype rule.
\item Fermentation rounds that draw unlabeled examples into the exchange, with a proof that the damping caused by wrong assignments is self-limiting for any assignment rule.
\item An evaluation on 14 datasets, 8 encoders and 310 task configurations with fixed constants, against baselines whose implementations were verified against reference code. KF is on par with the best tuned inductive baseline. With unlabeled carriers it gains 13.4 points at one shot and has the highest average accuracy among nine transductive methods, with a significant lead on vision encoders, and its geometry improves other transductive methods. Controlled studies on synthetic and real data attribute the gain to the geometry and tie it to the degree to which classes share their within-class variation.
\end{itemize}

\section{Related work}
\label{sec:related}

Few-shot classification with a fixed representation has turned out to be a strong paradigm. Prototypical networks classify by distance to class means \cite{snell2017prototypical}, and a linear classifier fitted on a good embedding matches far more elaborate meta-learners \cite{tian2020rethinking}. Large pretrained encoders, whether supervised \cite{he2016resnet,liu2022convnext,dosovitskiy2020vit}, self-supervised \cite{oquab2023dinov2}, language-supervised \cite{radford2021clip} or trained on text pairs \cite{reimers2019sbert,xiao2024cpack}, have made this setting the practical default, and they shift the difficulty from learning features to using a few of them well. Both the prototype rule and the linear probe ignore the anisotropy of within-class variation. Discriminant analysis accounts for it but depends on a covariance estimate, and in the regime of interest the dimension exceeds the number of residuals, so the estimate must be regularized, typically by linear shrinkage toward the identity \cite{ledoit2004well}.

A second line of work enlarges the training set in feature space. Hallucination approaches learn how base-class examples vary and transplant that variation to novel classes \cite{hariharan2017lowshot}, and distribution calibration borrows the mean and covariance of the nearest base classes and samples features from the calibrated Gaussian \cite{yang2022bridging}. Interpolation between training points regularizes without any statistics at all \cite{zhang2017mixup}. Implicit semantic data augmentation perturbs deep features along class-conditional covariance directions and, instead of sampling, minimizes a closed-form upper bound of the expected loss \cite{wang2021isda}; the general link between training with noise and regularization is classical \cite{bishop1995training}. KF shares the implicit treatment of noise but changes what the noise is made of. Its covariance is pooled across the classes of the task, each direction is certified on held-out classes, and its strength is set by a guarantee rather than by a schedule.

The statistical ingredient closest to ours is the estimation of covariance eigenvalues by sample splitting, where eigenvectors are computed on one part of the data and variances measured on another \cite{lam2016nercome}. We split by class rather than by sample, which is what makes the estimate meaningful with two examples per class, and we use the result one-sidedly: directions above the medium level are damped and nothing is ever amplified.

When unlabeled queries are available, transductive methods exploit them. Soft $k$-means refines prototypes with soft assignments of unlabeled points \cite{ren2018meta}, prototype rectification removes the bias between support and query features and adds confident pseudo-labeled queries \cite{liu2020prototype}, Laplacian regularization encourages neighboring queries to share a label \cite{ziko2020laplacian}, information maximization adapts a classifier by trading conditional against marginal entropy of the predictions \cite{boudiaf2020tim}, and optimal transport balances the assignments across classes \cite{hu2021leveraging}. Later methods weaken the assumption of balanced queries or refine the prototypes together with the graph that links them to the queries. Tsallis entropies relax the balance prior of information maximization \cite{veilleux2022realistic}, a description-length penalty lets the set of occupied classes adapt to the queries \cite{martin2022paddle}, label propagation over an iteratively refined graph is driven by the prototypes \cite{zhu2023protolp}, and conditional transport between prototypes and queries balances the uniform class prior of optimal transport against one adapted to the task \cite{tian2023putm}. All of them optimize prototypes, assignments or a transport cost for the queries at hand. Fermentation rounds instead re-estimate the within-class geometry from the carriers in closed form, which is where most of the gain in our experiments comes from.

Beyond few-shot classification, several recent methods let knowledge that a learning system already holds take the place of data that are scarce, costly to process or out of reach. Model annealing targets small labeled sets, which a deep network memorizes long before training ends, and spends the remaining budget on relearning from the network itself: from the temperature-softened predictions of a memory of its earlier states, and from adversarial inputs on which network and memory disagree most \cite{mohammadiruzbahani2026model}. Large language models have been used as meta-learners that operate on weights. Given the weight and gradient history of a training run, such a model writes updates that replace some of the passes over the training data, each kept only if it passes a check on held-out images \cite{ruzbahani2026gradreason}; given the client models in federated learning, it writes the aggregation rule, with no training on the server \cite{ruzbahani2026fedreason}. Where no example of an attack is available, a language-model agent composes an anomaly detector for a cyber-physical system from physical residuals, each fitted on one part of the normal data and tested on another \cite{ruzbahani2026pia}. These methods reuse the predictions and weights of networks in training or the physical priors of a language model. KF keeps the encoder frozen and reuses, from a few examples per class, the variation that the classes of one task share. It applies the same safeguard of held-out testing, with classes rather than examples held out.

\section{Knowledge fermentation}
\label{sec:method}

Let $f$ be a frozen encoder with outputs in $\R^d$. A task consists of $K$ classes with $n$ labeled examples each, the support set $\{(x_i,y_i)\}_{i=1}^{N}$ with $N=Kn$, and possibly a set of $M$ unlabeled examples $\{u_j\}_{j=1}^{M}$ drawn from the same classes. Every feature is scaled to unit length and centered at the mean of the features available to the learner, and $x_i,u_j\in\R^d$ denote the results. The \emph{strain} of class $c$ is its prototype $\hat{\mu}_c=\frac{1}{n}\sum_{i:y_i=c}x_i$. The \emph{carriers} are the residuals $r_i=x_i-\hat{\mu}_{y_i}$, stacked as the rows of $R\in\R^{N\times d}$; they carry $\nu=K(n-1)$ degrees of freedom and define the pooled scatter $S=R^{\T}R/\nu$. In established terms, strains are class prototypes, carriers are within-class residuals, the cross-fermentation of Section~\ref{sec:cross} is a cross-fitted estimate of the pooled within-class spectrum, and the rounds of Section~\ref{sec:rounds} are a transductive re-estimation of it.

The analysis rests on a single modeling assumption, the fermentation model
\begin{equation}
x=\mu_y+\xi,\qquad \E[\xi\mid y]=0,\qquad \operatorname{Cov}[\xi\mid y]=\bS ,
\label{eq:model}
\end{equation}
in which the nuisance covariance $\bS$ is shared by all classes and the nuisance terms of different examples are independent. Under \eqref{eq:model} the carriers of every class are observations of the same $\bS$, so knowledge about variation can be exchanged between classes. The question that organizes the method is which part of $\bS$ can be trusted when it must be learned from the carriers at hand.

\subsection{Cross-fermentation and the matured geometry}
\label{sec:cross}

The classes are split at random into $F$ groups $G_1,\dots,G_F$. For a group $G_f$, let $R_f$ collect the carriers of its classes, $R_{-f}$ the carriers of all other classes, and $\nu_f=|G_f|(n-1)$. Let $v^{(-f)}_1,v^{(-f)}_2,\dots$ be the orthonormal eigenvectors of $R_{-f}^{\T}R_{-f}$ in decreasing order of eigenvalue. The variance that the $i$-th of these directions carries on the held-out group is
\begin{equation}
h^{(f)}_i=\frac{1}{\nu_f}\bigl\|R_f\,v^{(-f)}_i\bigr\|^2 ,
\label{eq:heldout}
\end{equation}
and the \emph{matured spectrum} is the average $h_i=\frac{1}{F}\sum_{f}h^{(f)}_i$ for $i=1,\dots,r$, where $r$ is the smallest rank among the matrices $R_{-f}$. The value $h_i$ is attached to the $i$-th principal direction $v_i$ of the full scatter $S$, which is estimated from all carriers; Proposition~\ref{prop:spectrum} bounds what this transfer can cost. If $r<d$, the directions outside the span of $v_1,\dots,v_r$ have not been observed, and the variance left for each of them is $\rho=(\tr S-\sum_{i\le r}h_i)_+/(d-r)$. The \emph{medium level} is the median variance over all $d$ directions,
\begin{equation}
\hat{m}=\med\bigl\{h_1,\dots,h_r,\underbrace{\rho,\dots,\rho}_{d-r}\bigr\},
\label{eq:medium}
\end{equation}
which is the median of $h_1,\dots,h_d$ when $r=d$. A direction is \emph{mature} when $h_i>\hat{m}$. Writing $I$ for the set of mature directions, the structured noise and the fermented covariance are
\begin{equation}
\widehat{N}=\sum_{i\in I}(h_i-\hat{m})\,v_iv_i^{\T},\qquad \hbS=\hat{m}\,I_d+\widehat{N}.
\label{eq:noise}
\end{equation}
Only directions whose excess variance has been confirmed by classes that did not define them enter $\widehat{N}$. Everything else is treated as an isotropic medium. In the degenerate case $\hat{m}=0$ no direction is damped.

Structured noise could be injected by sampling, but its effect has a closed form. Noise with covariance proportional to $\hbS$ on the encoder features is isotropic noise after the linear map
\begin{equation}
A_s=I_d-\sum_{i\in I}\bigl(1-g_i^{\,s}\bigr)v_iv_i^{\T},\qquad g_i=\sqrt{\hat{m}/h_i},
\label{eq:map}
\end{equation}
taken at level $s=1$, and for Gaussian class conditionals with covariance $\hbS$ the optimal classifier is the nearest-prototype rule after that map (Proposition~\ref{prop:geometry}). The map damps each mature direction to the medium level and leaves all other directions untouched. Prototypes estimated from few examples have unreliable norms after any such map (Proposition~\ref{prop:energy}), so features are renormalized, which gives the \emph{fermented geometry}
\begin{equation}
\psi_s(x)=\frac{A_sx}{\|A_sx\|}.
\label{eq:psi}
\end{equation}
The exact statements above concern the linear map. On the sphere, $\psi_s$ measures similarity by the cosine in the inner product of $\hbS^{-s}$, a separate step whose properties are stated in Proposition~\ref{prop:geometry}.

The level $s\in\{0,\tfrac12,1\}$ is chosen by an \emph{integrity gate}. For each group $G_f$ the geometry is fermented from the classes outside $G_f$ alone. Every support example of a class in $G_f$ is then assigned to the nearest prototype in that geometry, with its own prototype recomputed without it. The gate returns the largest level that attains the highest fraction of correct assignments. It therefore admits structured noise at full strength unless there is evidence on held-out classes that damping removes identity, in which case it backs off.

\subsection{Integrity-calibrated training}
\label{sec:product}

In the fermented geometry the remaining within-class variation is close to isotropic, and the classifier is trained under isotropic noise. Let $\psi_i=\psi_s(x_i)$, let $p_c=\frac1n\sum_{i:y_i=c}\psi_i$, and let $D_c=\min_{j\ne c}\|p_c-p_j\|$ be the distance from strain $c$ to its nearest neighbor. The noise strength of class $c$ is
\begin{equation}
\lambda_c=\frac{D_c^2}{4z^2},
\label{eq:strength}
\end{equation}
where $z$ is the \emph{integrity level}. With this choice, noise of variance $\lambda_c$ added to $p_c$ moves it closer to any given other strain with probability at most $\bar{\Phi}(z)$, the standard normal tail, and the probability that it ends closer to some other strain is bounded by a sum over competitors whose terms decay rapidly with their distance (Theorem~\ref{thm:integrity}). We use $z=2$ throughout. A level that grows with the number of classes, $z=\bar{\Phi}^{-1}(\alpha/(K-1))$, bounds the second probability by $\alpha$; Section~\ref{sec:analysis} examines this choice.

For a linear classifier with weights $w_1,\dots,w_K$ and intercepts $b_1,\dots,b_K$, the expected cross-entropy under noise $\delta\sim\mathcal{N}(0,\lambda_{y_i}I_d)$ added to $\psi_i$ is bounded by the \emph{fermented risk}
\begin{equation}
\begin{split}
\mathcal{L}(W,b)=\frac1N\sum_{i=1}^{N}\log\sum_{j=1}^{K}\exp\Bigl(&\langle w_j-w_{y_i},\psi_i\rangle+b_j-b_{y_i}\\
&+\tfrac{\lambda_{y_i}}{2}\|w_j-w_{y_i}\|^2\Bigr).
\end{split}
\label{eq:risk}
\end{equation}
The bound follows the argument of implicit semantic augmentation \cite{wang2021isda}; here it is applied in a geometry where the noise is isotropic, with a strength that differs across classes. The fermented risk is convex with an essentially unique minimizer, and it connects the empirical risk at $\lambda=0$ with the inner-product prototype rule in the limit of large $\lambda$ (Theorem~\ref{thm:risk}). The integrity rule selects the operating point on this path without validation data. With a single example per class and no unlabeled carriers there is no within-class knowledge to exchange: no carrier exists, the gate has nothing to test, the geometry stays Euclidean, and the product is the cosine prototype rule.

\subsection{Fermentation rounds}
\label{sec:rounds}

Unlabeled examples carry the same nuisance variation as labeled ones, and there are usually many more of them. They are drawn into the exchange as follows. Let $A^{(t)}$ be the map of round $t$ and $\psi^{(t)}$ the corresponding geometry, with $A^{(0)}$ obtained from the labeled carriers as in Section~\ref{sec:cross}, or equal to the identity when $n=1$. Prototypes start at the labeled strains and are refined by three passes of
\begin{align}
\pi_{jc}&=\frac{\exp\bigl(\langle\psi^{(t)}(u_j),\bar{p}_c\rangle/\lambda^{(t)}\bigr)}{\sum_{c'}\exp\bigl(\langle\psi^{(t)}(u_j),\bar{p}_{c'}\rangle/\lambda^{(t)}\bigr)},\label{eq:post}\\
p_c&=\frac{\sum_{i:y_i=c}\psi^{(t)}(x_i)+\sum_j\pi_{jc}\,\psi^{(t)}(u_j)}{n+\sum_j\pi_{jc}},\label{eq:proto}
\end{align}
where $\bar{p}_c=p_c/\|p_c\|$ and $\lambda^{(t)}$ is the median over classes of the strength \eqref{eq:strength} computed from the normalized strains. The weights \eqref{eq:post} are soft assignments. They equal the posterior probability that $u_j$ is a noisy copy of strain $c$ when all classes are equally likely and the noise is isotropic with the common variance $\lambda^{(t)}$, so the same constant $z$ governs training and assignment, and no class proportions are estimated.

The geometry is then matured. Class means are recomputed in the original feature space with the weights $\pi_{jc}$, each unlabeled example becomes a carrier $u_j-\mu_{\hat{c}_j}$ of its most probable class $\hat{c}_j$, and cross-fermentation is applied to the $N+M$ carriers, each unlabeled carrier following the group of its assigned class. This yields $A^{(t+1)}$ at full level. After $T$ rounds the labeled strains are used once more to minimize the fermented risk \eqref{eq:risk} in the final geometry, and the prediction for $u$ multiplies the two sources of evidence,
\begin{equation}
q(c\mid u)\;\propto\;\pi_c(u)\,\frac{\exp(\langle w_c,\psi^{(T)}(u)\rangle+b_c)}{\sum_{c'}\exp(\langle w_{c'},\psi^{(T)}(u)\rangle+b_{c'})}.
\label{eq:fusion}
\end{equation}
The first factor is what the matured population of carriers says about $u$; the second is what the labeled strains say. Neither factor constrains how the predictions are distributed over the classes, in contrast to rules that balance their predictions across the unlabeled set. Algorithm~\ref{alg:kf} lists the complete procedure. We write KF for the form that uses labeled carriers only and KF-T for the form with rounds. The final geometry $\psi^{(T)}$ is also useful on its own: any other transductive rule can be run on $\psi^{(T)}(x_i)$ and $\psi^{(T)}(u_j)$ instead of the encoder features (Section~\ref{sec:results}).

\begin{algorithm}[t]
\caption{Knowledge Fermentation}
\label{alg:kf}
\begin{algorithmic}[1]
\Require strains and carriers of the support set; unlabeled set (optional); $F$, $z$, $T$
\State split the classes into $F$ groups; compute the matured spectrum \eqref{eq:heldout} and the medium level \eqref{eq:medium}
\State form the map $A_s$ \eqref{eq:map}; select the level $s$ with the integrity gate
\For{$t=0,\dots,T-1$ \textbf{if} unlabeled examples exist}
\State refine prototypes and assignments with \eqref{eq:post} and \eqref{eq:proto}
\State add unlabeled carriers; repeat step 1 on all carriers to obtain $A^{(t+1)}$
\EndFor
\State \textbf{if} unlabeled examples exist, refine prototypes and assignments in $\psi^{(T)}$
\State set the strengths \eqref{eq:strength}; minimize the fermented risk \eqref{eq:risk}
\State predict with the linear rule, or with the fusion \eqref{eq:fusion} when unlabeled examples exist
\end{algorithmic}
\end{algorithm}

The cost is dominated by $(F+1)^2$ eigendecompositions of size at most $\min(N,d)$ for cross-fermentation with its gate, $F+1$ of size $d$ per round, and a smooth convex problem in $Kd$ variables. There is no sampling and no stochastic optimization.

\section{Theoretical analysis}
\label{sec:theory}

This section states the properties that justify each step. Proofs are given in \ref{app:proofs}. All statements refer to the fermentation model \eqref{eq:model}; Gaussianity of $\xi$ is assumed only where stated.

The first result explains why the spectrum has to be measured on held-out classes.

\begin{proposition}[Held-out and in-sample spectra]
\label{prop:spectrum}
Let $u$ be a unit vector that depends only on the carriers outside $G_f$. Then
\begin{equation}
\E\Bigl[\tfrac{1}{\nu_f}\|R_fu\|^2\,\Big|\,R_{-f}\Bigr]=u^{\T}\bS u ,
\label{eq:unbiased}
\end{equation}
and if $\xi$ is Gaussian the conditional variance equals $2(u^{\T}\bS u)^2/\nu_f$. Let $\ell_1\ge\dots\ge\ell_d$ be the eigenvalues of $\bS$. For unit vectors $u$ and $v$ whose spans enclose the angle $\theta$,
\begin{equation}
\bigl|v^{\T}\bS v-u^{\T}\bS u\bigr|\;\le\;(\ell_1-\ell_d)\sin\theta .
\label{eq:transfer}
\end{equation}
In contrast, let $\hat{\ell}_1\ge\hat{\ell}_2\ge\dots$ be the eigenvalues of $S$, with eigenvectors $v_1,v_2,\dots$ as in Section~\ref{sec:cross}. For every $k$,
\begin{equation}
\E\Bigl[\sum_{i\le k}\hat{\ell}_i\Bigr]\;\ge\;\sum_{i\le k}\ell_i\;\ge\;\E\Bigl[\sum_{i\le k}v_i^{\T}\bS\,v_i\Bigr].
\label{eq:insample}
\end{equation}
When $\nu<d$, the gap between the two outer terms at $k=\nu$ is at least $\sum_{i>\nu}\ell_i$.
\end{proposition}

The in-sample eigenvalues therefore overstate the variance that the estimated directions really carry, and in the few-shot regime they attribute to $\nu$ directions the variance of all $d$. The held-out value \eqref{eq:heldout} measures without bias the variance of the direction on which it is taken, with a relative error of order $\sqrt{2/\nu_f}$ in the Gaussian case. That direction, $v^{(-f)}_i$, is estimated without the classes of one group, whereas the map \eqref{eq:map} acts on $v_i$, estimated from all classes. By \eqref{eq:transfer} the two variances differ by at most $(\ell_1-\ell_d)\sin\theta^{(f)}_i$, where $\theta^{(f)}_i$ is the angle between the two directions, so that
\begin{equation}
\bigl|\E\,h_i-\E\,v_i^{\T}\bS v_i\bigr|\;\le\;(\ell_1-\ell_d)\,\E\Bigl[\tfrac1F\sum_f\sin\theta^{(f)}_i\Bigr].
\label{eq:transferbias}
\end{equation}
The matured spectrum is thus unbiased for the directions of the folds, and its error for the directions of the full scatter is controlled by how far a principal direction moves when one group of classes is removed. Section~\ref{sec:analysis} measures both terms with a known covariance and on the benchmark. The argument behind \eqref{eq:unbiased} also shows that $\rho$, before truncation at zero, is unbiased for the average variance outside the directions on which the $h_i$ were measured, which is what the medium level \eqref{eq:medium} requires.

\begin{proposition}[Fermented geometry]
\label{prop:geometry}
Let $\hbS$, $A_s$ and $\psi_s$ be defined by \eqref{eq:noise}, \eqref{eq:map} and \eqref{eq:psi} with $\hat{m}>0$, and let $g_{\min}=\min_{i\in I}g_i$, with $g_{\min}=1$ if $I$ is empty. Then
(a) $A_s=(\hbS/\hat{m})^{-s/2}$ for $s\in[0,1]$, hence $A_1\hbS A_1^{\T}=\hat{m}I_d$;
(b) if $\epsilon\sim\mathcal{N}(0,(\lambda/\hat{m})\hbS)$, then $A_1\epsilon\sim\mathcal{N}(0,\lambda I_d)$ and $\langle w,x+\epsilon\rangle=\langle A_1^{-1}w,A_1x+A_1\epsilon\rangle$ for all $w$ and $x$, so the loss of a linear classifier $(W,b)$ under the structured noise $\epsilon$ on $x$ has the distribution of the loss of $(WA_1^{-1},b)$ under isotropic noise of variance $\lambda$ on $A_1x$;
(c) for class conditionals $\mathcal{N}(\mu_c,\hbS)$ with equal priors the Bayes rule is $\argmin_c\|A_1(x-\mu_c)\|$;
(d) for all $s\in[0,1]$ and all $x,x'$, $A_1\preceq A_s\preceq I_d$, $\|A_sx-A_sx'\|\le\|x-x'\|$ and $\|A_sx\|\ge g_{\min}^{\,s}\|x\|$;
(e) for nonzero $x,x'$,
\begin{equation}
\langle\psi_s(x),\psi_s(x')\rangle=\frac{x^{\T}\hbS^{-s}x'}{\sqrt{x^{\T}\hbS^{-s}x}\,\sqrt{x'^{\T}\hbS^{-s}x'}}
\label{eq:cosine}
\end{equation}
and $\|\psi_s(x)-\psi_s(x')\|\le g_{\min}^{-s}\,\bigl\|x/\|x\|-x'/\|x'\|\bigr\|$.
\end{proposition}

Parts (b) and (c) are the precise sense in which the matured noise amounts to a change of geometry, and both concern the linear map. Training a linear classifier under structured noise on the encoder features is the same problem as training it under isotropic noise after $A_1$, and under a Gaussian model with the matured covariance the optimal rule is the prototype rule after $A_1$. Part (d) states that the map is one-sided: it leaves untested directions untouched and contracts a tested direction by no more than its measured ratio to the medium. Plug-in whitening with a shrunk scatter matrix, by comparison, contracts the whole observed subspace relative to the unobserved one, without bound as the shrinkage vanishes. Part (e) describes the geometry the algorithm works in. After normalization, similarity is the cosine in the inner product of $\hbS^{-s}$; the Gaussian model of (c) no longer holds exactly, and distances between normalized points can grow, but by at most the factor $g_{\min}^{-s}=(\max_ih_i/\hat{m})^{s/2}$.

\begin{proposition}[Strain energy]
\label{prop:energy}
Assume $\xi$ Gaussian and $\hbS=\bS$ with medium level $m$. For the estimated strain $\hat{\mu}_c=\mu_c+\bar{\xi}_c$,
\begin{align}
\E\|A_1\hat{\mu}_c\|^2&=\|A_1\mu_c\|^2+\frac{md}{n},\\
\operatorname{Var}\|A_1\hat{\mu}_c\|^2&=\frac{2m^2d}{n^2}+\frac{4m\|A_1\mu_c\|^2}{n}.
\end{align}
\end{proposition}

With estimated strains, the Bayes rule of Proposition~\ref{prop:geometry}(c) contains the offset $-\|A_1\hat\mu_c\|^2/(2m)$. By Proposition~\ref{prop:energy} the difference of two such offsets has variance at least $d/n^2$, while the quantity it competes with, half the squared Mahalanobis distance between two classes, does not grow with $d$. For $d$ in the hundreds and $n$ in the single digits the offsets are mostly noise. Normalizing features and strains as in \eqref{eq:psi} removes them, which is why the fermented geometry is defined on the sphere.

\begin{theorem}[Integrity]
\label{thm:integrity}
Let $p_1,\dots,p_K$ be the strains in the fermented geometry, $D_{cj}=\|p_c-p_j\|$, $D_c=\min_{j\ne c}D_{cj}$, and $\delta\sim\mathcal{N}(0,\lambda_cI_d)$ with $\lambda_c=D_c^2/(4z^2)$. Then for every $j\ne c$
\begin{equation}
\Prob\bigl(\|p_c+\delta-p_j\|\le\|\delta\|\bigr)=\bar{\Phi}\bigl(zD_{cj}/D_c\bigr)\le\bar{\Phi}(z),
\end{equation}
and the probability that $p_c+\delta$ is closer to some other strain than to $p_c$ is at most
\begin{equation}
\sum_{j\ne c}\bar{\Phi}\bigl(zD_{cj}/D_c\bigr)\;\le\;(K-1)\,\bar{\Phi}(z),
\label{eq:union}
\end{equation}
which does not exceed $\alpha$ when $z\ge\bar{\Phi}^{-1}(\alpha/(K-1))$. More generally, a point whose distance to the bisector of $p_c$ and $p_j$ is $a$, on the side of $p_c$, crosses it with probability $\bar{\Phi}(a/\sqrt{\lambda_c})$. At level $s=1$ and before normalization, $\delta$ corresponds in the encoder space to noise with covariance $(\lambda_c/\hat{m})\,\hbS$.
\end{theorem}

At a fixed level the guarantee is pairwise: toward any single competitor a fermented strain crosses with probability at most $\bar{\Phi}(z)$, which is $2.3\%$ at $z=2$. The probability of leaving the class altogether is governed by the sum in \eqref{eq:union}. A competitor at twice the nearest distance contributes $\bar{\Phi}(2z)$, below $10^{-4}$ at $z=2$, so the sum is determined by the number of strains near the nearest distance rather than by $K$ itself; a level that grows with $K$ as stated caps it at $\alpha$. The statement about a general point covers support examples, which lie at other distances from the bisectors than the strains do. The last statement closes the loop with the starting idea: isotropic noise in the fermented geometry is structured noise on the encoder features, shaped by the fermented covariance, and the integrity level fixes how much of it a class can absorb.

\begin{theorem}[Fermented risk]
\label{thm:risk}
Let $\mathcal{L}_0$ be the empirical cross-entropy, $\mathcal{L}_\infty$ its expectation under $\delta\sim\mathcal{N}(0,\lambda_{y_i}I_d)$ added to $\psi_i$, and $\mathcal{L}$ the fermented risk \eqref{eq:risk}. Then (a) $\mathcal{L}_0\le\mathcal{L}_\infty\le\mathcal{L}$ and $\mathcal{L}-\mathcal{L}_\infty\le\max_{i,j}\tfrac{\lambda_{y_i}}{2}\|w_j-w_{y_i}\|^2$; (b) $\mathcal{L}$ is convex; if $\lambda_c>0$ for every class it attains its minimum, and any two minimizers differ only by a common shift of all $w_c$ and a common shift of all $b_c$, which do not change the classifier; (c) for $\lambda_c\equiv\lambda$, classes of equal size and no intercepts, the minimizer $W_\lambda$ with $\sum_cw_c=0$ satisfies $\lambda W_\lambda\to\tfrac12(P-\mathbf{1}\bar{p}^{\T})$ as $\lambda\to\infty$, where the rows of $P$ are the strains $p_c$ and $\bar{p}$ is their mean; consequently, at every $x$ for which $\max_c\langle p_c,x\rangle$ is attained by a single class, the prediction of $W_\lambda$ equals $\argmax_c\langle p_c,x\rangle$ for all sufficiently large $\lambda$, while $\mathcal{L}\to\mathcal{L}_0$ pointwise as $\lambda\to0$.
\end{theorem}

Part (c) places two classical few-shot classifiers at the ends of a single path. Weak fermentation recovers the linear probe. Strong fermentation recovers the inner-product prototype rule $\argmax_c\langle p_c,x\rangle$, which coincides with the nearest-prototype rule $\argmax_c(\langle p_c,x\rangle-\tfrac12\|p_c\|^2)$ whenever the strains have equal norm, as they do exactly for $n=1$, where they lie on the sphere. The integrity rule \eqref{eq:strength} picks a point in between that depends only on the geometry of the strains.

The last result concerns the rounds, in which carriers are formed with assignments that may be wrong.

\begin{proposition}[Self-limiting maturation]
\label{prop:rounds}
Let $(x,y)$ follow \eqref{eq:model}, let the class $a$ attributed to $x$ be any random variable, possibly a function of $x$, with $q_{cc'}=\Prob(y=c,a=c')$, and let the carrier of $x$ be $x-m_a$ for fixed centers $m_1,\dots,m_K$. For a unit vector $v$ put $\sigma_v^2=v^{\T}\bS v$, $\Delta_v=\max_{c\ne c'}|v^{\T}(\mu_c-\mu_{c'})|$, $\eta_v=\max_c|v^{\T}(\mu_c-m_c)|$ and $\varepsilon=\sum_{c\ne c'}q_{cc'}$, and let $\tau_v^2$ be the second moment of the carriers along $v$. Then
\begin{equation}
\tau_v\;\le\;\sigma_v+\sqrt{\varepsilon}\,\Delta_v+\eta_v ,
\label{eq:leak}
\end{equation}
and if direction $v$ is damped with gain $g=\min\{1,\sqrt{m}/\tau_v\}$, the largest class separation along $v$ after damping satisfies
\begin{equation}
g\,\Delta_v\;\ge\;\min\Bigl\{\Delta_v,\;\sqrt{m}\,\Bigl(\frac{\sigma_v+\eta_v}{\Delta_v}+\sqrt{\varepsilon}\Bigr)^{-1}\Bigr\}.
\label{eq:selflimit}
\end{equation}
If, in addition, $a$ is independent of $\xi$ given $y$ and $m_c=\mu_c$, the second moment of the carriers is $\bS+B_q$ with $B_q=\sum_{c\ne c'}q_{cc'}(\mu_c-\mu_{c'})(\mu_c-\mu_{c'})^{\T}$, and the second term of the minimum improves to $\sqrt{m}\,(\sigma_v^2/\Delta_v^2+\varepsilon)^{-1/2}$.
\end{proposition}

Wrong assignments leak between-class variance into the carriers, and the leak triggers damping along discriminative directions. The proposition shows that this effect limits itself, and it does so for any assignment rule: the attribution may depend on the features, as the soft assignments \eqref{eq:post} do, and the centers may be estimates. The guaranteed separation increases with $\Delta_v$, and for a strongly discriminative direction, where $\Delta_v\gg\sigma_v+\eta_v$, it approaches $\sqrt{m/\varepsilon}$, the noise scale of the medium divided by the square root of the error rate. Dependence between assignment and nuisance costs at most a factor $\sqrt2$ in this bound when the centers are exact. As assignments improve over the rounds, $\varepsilon$ decreases and the bound approaches the separation after ideal whitening. The statement concerns population moments along a fixed direction; the finite-sample behavior of the rounds under controlled assignment errors is examined in Section~\ref{sec:analysis}.

\section{Experiments}
\label{sec:experiments}

The experiments address four questions. Does fermentation help when only labeled examples exist? How much do unlabeled carriers add, and how does the result compare with transductive methods that see the same data, also when the unlabeled set is small, imbalanced or incomplete? Which components produce the gain? Do the mechanisms behave as Section~\ref{sec:theory} predicts, under a known covariance and on the benchmark?

\begin{table*}[t]
\centering
\caption{Inductive accuracy (\%).}
\label{tab:inductive}
\small
\begin{tabular}{@{}lccccccccccccccc@{}}
\toprule
 & \multicolumn{5}{c}{Vision} & \multicolumn{5}{c}{Text} & \multicolumn{5}{c}{All} \\
\cmidrule(lr){2-6}\cmidrule(lr){7-11}\cmidrule(lr){12-16}
Method & 1 & 2 & 4 & 8 & 16 & 1 & 2 & 4 & 8 & 16 & 1 & 2 & 4 & 8 & 16 \\
\midrule
NCM & 51.9 & 61.3 & 68.6 & 73.2 & 76.3 & \underline{54.6} & 66.1 & 74.0 & 79.2 & 82.8 & 52.4 & 62.2 & 69.6 & 74.3 & 77.5 \\
LR & \underline{52.2} & 62.9 & 71.0 & 76.8 & 81.2 & \textbf{54.7} & 66.3 & 74.5 & \underline{80.3} & \underline{84.1} & \underline{52.7} & 63.5 & 71.7 & 77.5 & 81.8 \\
LDA & 51.9 & 62.6 & 71.1 & 77.1 & 81.5 & \underline{54.6} & 64.5 & 73.2 & 79.5 & 83.6 & 52.4 & 62.9 & 71.5 & 77.6 & 81.9 \\
Noise & 51.0 & 61.9 & 70.4 & 76.5 & 81.1 & 53.8 & 65.6 & 73.9 & 79.9 & 84.0 & 51.5 & 62.6 & 71.1 & 77.2 & 81.7 \\
Mixup & 51.3 & 62.3 & 70.9 & 77.0 & 81.3 & 53.9 & 65.9 & 74.0 & 80.0 & 83.9 & 51.8 & 63.0 & 71.5 & 77.6 & 81.8 \\
ISDA & \textbf{52.3} & \underline{63.0} & \underline{71.8} & \underline{77.7} & \underline{82.0} & \underline{54.6} & \textbf{66.5} & \underline{74.6} & \textbf{80.5} & \textbf{84.2} & \textbf{52.8} & \underline{63.7} & \underline{72.3} & \underline{78.2} & \underline{82.4} \\
\midrule
KF & 51.9 & \textbf{63.3} & \textbf{72.2} & \textbf{78.1} & \textbf{82.5} & \underline{54.6} & \underline{66.4} & \textbf{74.7} & \textbf{80.5} & \underline{84.1} & 52.4 & \textbf{63.9} & \textbf{72.7} & \textbf{78.6} & \textbf{82.8} \\
\bottomrule
\end{tabular}
\end{table*}

\begin{table*}[t]
\centering
\caption{Transductive accuracy (\%).}
\label{tab:transductive}
\small
\begin{tabular}{@{}lccccccccccccccc@{}}
\toprule
 & \multicolumn{5}{c}{Vision} & \multicolumn{5}{c}{Text} & \multicolumn{5}{c}{All} \\
\cmidrule(lr){2-6}\cmidrule(lr){7-11}\cmidrule(lr){12-16}
Method & 1 & 2 & 4 & 8 & 16 & 1 & 2 & 4 & 8 & 16 & 1 & 2 & 4 & 8 & 16 \\
\midrule
Soft $k$-means & 58.8 & 65.4 & 70.6 & 73.4 & 75.3 & 70.6 & 77.0 & 79.7 & 82.0 & 83.7 & 61.1 & 67.7 & 72.4 & 75.1 & 76.9 \\
BD-CSPN & 58.0 & 65.8 & 70.2 & 73.1 & 75.2 & 64.6 & 72.9 & 77.0 & 79.7 & 82.4 & 59.3 & 67.2 & 71.5 & 74.4 & 76.6 \\
LaplacianShot & 56.7 & 64.0 & 69.8 & 74.0 & 76.8 & 64.4 & 73.6 & 75.1 & 80.4 & 83.4 & 58.2 & 65.9 & 70.8 & 75.3 & 78.1 \\
TIM & 59.9 & 67.6 & 72.1 & 75.4 & 77.9 & \textbf{73.8} & \textbf{81.1} & \textbf{83.7} & \textbf{85.6} & \textbf{86.9} & 62.6 & 70.2 & 74.3 & 77.4 & 79.7 \\
PT-MAP & 56.7 & 62.8 & 66.1 & 68.0 & 69.3 & \underline{72.3} & 79.1 & 81.8 & 83.1 & 84.1 & 59.7 & 66.0 & 69.1 & 70.9 & 72.2 \\
PT-MAP (power) & 58.0 & 64.0 & 67.1 & 68.9 & 70.1 & 71.6 & 79.3 & 81.6 & 82.9 & 83.9 & 60.6 & 67.0 & 69.9 & 71.6 & 72.8 \\
$\alpha$-TIM & \underline{61.7} & \underline{70.6} & \underline{75.8} & \underline{78.9} & \underline{82.8} & 69.7 & 78.4 & \underline{82.2} & 81.9 & \underline{85.9} & \underline{63.2} & \underline{72.1} & \underline{77.0} & \underline{79.4} & \underline{83.4} \\
PADDLE & 51.0 & 57.3 & 61.5 & 65.3 & 68.9 & 54.3 & 62.5 & 68.7 & 73.9 & 78.7 & 51.7 & 58.3 & 62.9 & 66.9 & 70.8 \\
protoLP & 56.6 & 62.6 & 65.3 & 67.1 & 68.4 & 72.2 & 78.9 & 80.8 & 81.9 & 83.1 & 59.6 & 65.7 & 68.3 & 70.0 & 71.3 \\
\midrule
KF-T & \textbf{64.3} & \textbf{73.9} & \textbf{78.9} & \textbf{82.1} & \textbf{84.2} & 71.9 & \underline{79.4} & \underline{82.2} & \underline{84.0} & 85.4 & \textbf{65.8} & \textbf{75.0} & \textbf{79.5} & \textbf{82.5} & \textbf{84.5} \\
\bottomrule
\end{tabular}
\end{table*}

\subsection{Setup}
\label{sec:setup}

The benchmark comprises 14 datasets. Ten are image collections: CIFAR-10 and CIFAR-100, two labeled subsets of the Tiny Images collection \cite{torralba2008tiny}, Flowers-102 \cite{nilsback2008flowers}, the Describable Textures Dataset (DTD) \cite{cimpoi2014dtd}, Oxford-IIIT Pets \cite{parkhi2012pets}, CUB-200-2011 \cite{wah2022cub}, the remote sensing sets EuroSAT \cite{helber2019eurosat} and RESISC45 \cite{cheng2017resisc}, and the medical sets BloodMNIST and PathMNIST from MedMNIST~v2 \cite{yang2023medmnist}. Four are text collections: the intent datasets Banking77 \cite{casanueva2020banking77} and CLINC150 \cite{larson2019clinc}, the latter without its out-of-scope queries, the fourteen-class ontology classification set derived from DBpedia \cite{lehmann2015dbpedia}, and the 20~Newsgroups collection attributed to Lang \cite{lang1995newsweeder}. The number of classes ranges from 8 to 200.

Images are embedded with five frozen encoders that differ in architecture and in pretraining signal: ResNet-50 \cite{he2016resnet} and ConvNeXt-T \cite{liu2022convnext} trained with labels, a supervised ViT-B/16 \cite{dosovitskiy2020vit}, the self-supervised DINOv2 ViT-S/14 \cite{oquab2023dinov2}, and the image tower of CLIP ViT-B/16 \cite{radford2021clip}. Text is embedded with three sentence encoders: MiniLM-L6 and MPNet-base from the Sentence-Transformers library, which follows the Sentence-BERT design \cite{reimers2019sbert}, and BGE-base \cite{xiao2024cpack}. Feature dimensions range from 384 to 2048; \ref{app:details} lists the checkpoints.

Every task uses all classes of its dataset. For each of the 62 dataset-encoder pairs and each $n\in\{1,2,4,8,16\}$, ten support sets are drawn from the training split and the classifier is evaluated on the test split, which gives 310 configurations and 3{,}100 episodes. Training pools are capped at 200 examples per class and test sets at 10,000 examples. In the transductive protocol the test features of a task are available without labels. Reported numbers are accuracies averaged over episodes and, in aggregate tables, over pairs. All methods share the support sets, so differences are paired. Their 95\% intervals come from a bootstrap that resamples datasets, which does not treat configurations of the same dataset as independent, and significance is assessed with a Wilcoxon signed-rank test over configurations and, more conservatively, over datasets (Table~\ref{tab:statistics}).

KF is compared with six inductive methods that see the same labeled features: the nearest class mean rule (NCM) \cite{snell2017prototypical}, logistic regression (LR) \cite{tian2020rethinking}, linear discriminant analysis with Ledoit-Wolf shrinkage (LDA) \cite{ledoit2004well}, training with isotropic Gaussian feature noise (Noise), mixup on features \cite{zhang2017mixup}, and ISDA \cite{wang2021isda} with class-wise covariances. KF-T is compared with eight transductive methods: soft $k$-means \cite{ren2018meta}, BD-CSPN \cite{liu2020prototype}, LaplacianShot \cite{ziko2020laplacian}, TIM \cite{boudiaf2020tim}, PT-MAP \cite{hu2021leveraging}, $\alpha$-TIM \cite{veilleux2022realistic}, PADDLE \cite{martin2022paddle} and protoLP \cite{zhu2023protolp}. PUTM \cite{tian2023putm} trains a network for every task and is some fifteen to twenty times as costly as TIM; it is included in the study of imbalanced unlabeled sets, the setting it was designed for. All baselines receive the same normalized and centered features, with three exceptions: protoLP, PUTM and one of two forms of PT-MAP apply the preprocessing of their reference code to the normalized features. That preprocessing contains a power transform, which presupposes non-negative features, whereas most encoders produce signed ones; a signed version of the transform is therefore used, and PT-MAP is also run without it. Soft $k$-means is given a temperature. Every implementation was checked against reference code on identical tasks (\ref{app:details}). The hyperparameters of every baseline are selected separately for each shot count on six development pairs, with held-out training examples serving as validation queries. All constants of KF are fixed in every experiment: $F=2$ class groups, integrity level $z=2$, gate levels $\{0,\tfrac12,1\}$ and, for KF-T, $T=3$ rounds of three refinement passes.

KF was developed on 33 of the 62 pairs, among them the six used to tune the baselines; the other 29 were first touched in the final run, and Section~\ref{sec:results} reports them separately. The controlled studies of Section~\ref{sec:analysis} use five support sets per configuration and subsets of pairs that cover both modalities: 11 pairs for the components, 6 for the constants and the size of the unlabeled set, and 8 for the composition of the unlabeled set.

\begin{figure*}[t]
\centering
\includegraphics[width=0.96\textwidth]{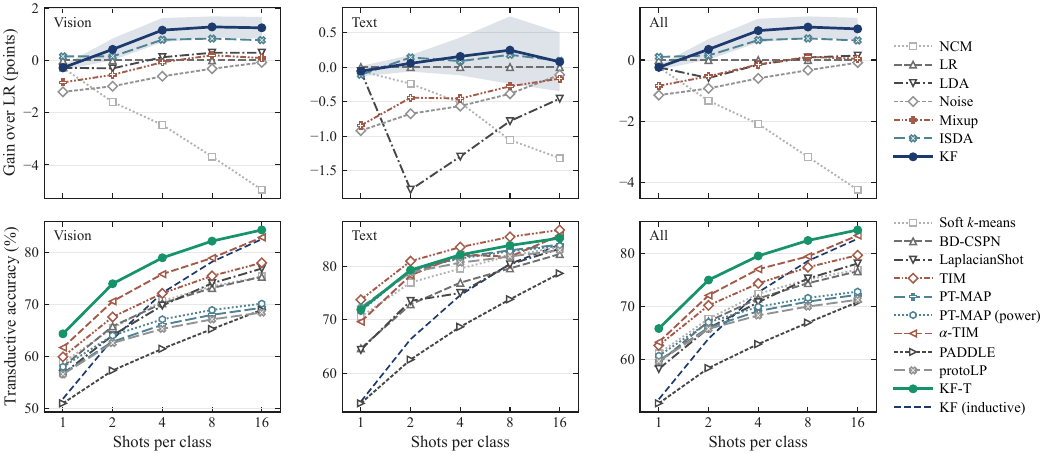}
\caption{Gain over LR (top) and transductive accuracy (bottom) versus shots.}
\label{fig:shots}
\end{figure*}

\begin{table}[t]
\centering
\caption{Transductive baselines on encoder features (Orig.) and in the fermented geometry (Ferm.); accuracy (\%) averaged over shots.}
\label{tab:plugin}
\footnotesize
\begin{tabular}{@{}lcccccc@{}}
\toprule
 & \multicolumn{2}{c}{Vision} & \multicolumn{2}{c}{Text} & \multicolumn{2}{c}{All} \\
\cmidrule(lr){2-3}\cmidrule(lr){4-5}\cmidrule(lr){6-7}
Method & Orig. & Ferm. & Orig. & Ferm. & Orig. & Ferm. \\
\midrule
Soft $k$-means & 68.7 & \textbf{75.9} & 78.6 & \textbf{80.6} & 70.6 & \textbf{76.8} \\
BD-CSPN & 68.5 & \textbf{76.1} & 75.3 & \textbf{79.3} & 69.8 & \textbf{76.7} \\
LaplacianShot & 68.3 & \textbf{74.2} & 75.4 & \textbf{78.1} & 69.6 & \textbf{74.9} \\
TIM & 70.6 & \textbf{74.0} & 82.2 & 82.2 & 72.8 & \textbf{75.6} \\
PT-MAP & 64.6 & \textbf{72.0} & 80.1 & \textbf{81.2} & 67.6 & \textbf{73.8} \\
\bottomrule
\end{tabular}
\end{table}

\begin{table*}[t]
\centering
\caption{Accuracy by encoder (\%), averaged over datasets and shots.}
\label{tab:encoder}
\small
\begin{tabular}{@{}lcccccccc@{}}
\toprule
 & \multicolumn{5}{c}{Vision encoders} & \multicolumn{3}{c}{Text encoders} \\
\cmidrule(lr){2-6}\cmidrule(lr){7-9}
 &  &  &  & DINOv2 & CLIP &  &  &  \\
Method & ResNet-50 & ConvNeXt-T & ViT-B/16 & ViT-S/14 & ViT-B/16 & MiniLM-L6 & MPNet-base & BGE-base \\
\midrule
NCM & 57.8 & 68.2 & 72.1 & 71.2 & 61.8 & 68.4 & 71.8 & 73.8 \\
LR & 60.0 & 70.5 & 74.7 & 73.1 & 65.7 & \underline{69.2} & \underline{72.5} & \textbf{74.3} \\
LDA & 59.6 & 70.5 & 75.1 & 72.7 & 66.3 & 68.3 & 71.9 & 73.1 \\
Noise & 59.5 & 69.9 & 74.1 & 72.7 & 64.8 & 68.6 & 71.9 & \underline{73.9} \\
Mixup & 59.8 & 70.2 & 74.5 & 72.9 & 65.5 & 68.7 & 72.0 & \underline{73.9} \\
ISDA & \underline{60.5} & \underline{71.0} & \underline{75.2} & \underline{73.4} & \underline{66.7} & \textbf{69.3} & \textbf{72.6} & \textbf{74.3} \\
KF & \textbf{60.7} & \textbf{71.1} & \textbf{75.3} & \textbf{73.5} & \textbf{67.3} & \textbf{69.3} & \textbf{72.6} & \textbf{74.3} \\
\midrule
Soft $k$-means & 60.7 & 70.2 & 74.5 & 74.0 & 64.2 & 75.4 & 79.1 & 81.3 \\
BD-CSPN & 60.0 & 70.2 & 74.1 & 73.6 & 64.5 & 72.2 & 75.6 & 78.2 \\
LaplacianShot & 59.6 & 70.2 & 74.3 & 73.6 & 63.6 & 72.3 & 75.9 & 77.9 \\
TIM & 62.9 & 71.9 & 76.3 & 74.2 & 67.6 & \textbf{79.3} & \textbf{82.8} & \textbf{84.5} \\
PT-MAP & 57.3 & 66.2 & 69.9 & 69.4 & 60.2 & 76.8 & 80.8 & \underline{82.6} \\
PT-MAP (power) & 59.2 & 66.9 & 72.0 & 69.2 & 60.8 & 76.6 & 80.7 & 82.3 \\
$\alpha$-TIM & \underline{65.6} & \underline{75.3} & \underline{79.9} & \underline{77.9} & \underline{71.0} & 76.7 & 80.5 & 81.7 \\
PADDLE & 57.2 & 61.6 & 68.3 & 64.2 & 52.8 & 64.7 & 70.1 & 68.1 \\
protoLP & 57.6 & 65.3 & 70.4 & 67.6 & 59.1 & 75.8 & 80.3 & 82.0 \\
KF-T & \textbf{66.8} & \textbf{78.4} & \textbf{82.6} & \textbf{80.2} & \textbf{75.5} & \underline{77.6} & \underline{81.9} & 82.2 \\
\bottomrule
\end{tabular}
\end{table*}

\begin{figure*}[t]
\centering
\includegraphics[width=0.9\textwidth]{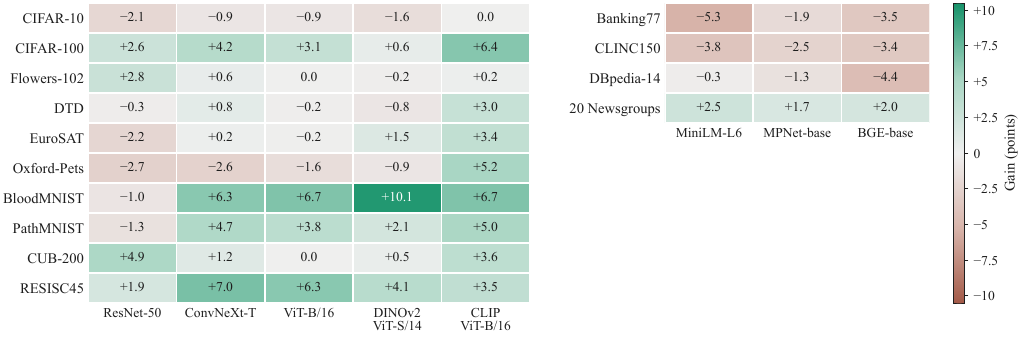}
\caption{Gain of KF-T over the best transductive baseline (points), averaged over shots.}
\label{fig:heatmap}
\end{figure*}

\subsection{Results}
\label{sec:results}

Table~\ref{tab:inductive} answers the first question. From two shots upward KF has the highest average accuracy over the whole benchmark. Over the 248 configurations with at least two shots it improves on NCM by 3.56 points, on Noise by 1.34, on Mixup by 0.99, on LDA by 0.98 and on LR by 0.85, and each of these differences remains significant when datasets rather than configurations are the unit of the test ($p\le0.005$). ISDA, whose strength and weight decay are tuned for every shot count, is the closest competitor. KF exceeds it by 0.32 points with fixed constants, with a 95\% interval of $[{-}0.06,\,0.69]$, and differs from the best baseline of each configuration by 0.07 points $[{-}0.26,\,0.40]$: without tuning, KF is on par with the best tuned inductive method. At one shot no within-class carrier exists and KF equals the prototype rule. The size of the gain follows the encoder. On CLIP features, whose within-class scatter is dominated by a few shared directions, KF exceeds LR by 2.12 points. On the sentence encoders that scatter is nearly isotropic, the integrity gate leaves the geometry almost unchanged, and the best tuned baseline of each configuration is ahead of KF by 0.37 points.

Unlabeled carriers change the magnitude of the effect (Table~\ref{tab:transductive}, Figure~\ref{fig:shots}). With one labeled example per class, KF-T reaches 65.8\% against 52.4\% for KF, a gain of 13.4 points; the gain is 11.1 points at two shots, 6.9 at four, and still 1.7 at sixteen. Averaged over the benchmark, KF-T is the most accurate of the nine transductive methods at every shot count. The strongest competitor on average is $\alpha$-TIM, which KF-T exceeds by 2.4 points, and KF-T ranks first in 181 of the 310 configurations. Its advantage is significant over datasets against every baseline of Table~\ref{tab:transductive} except TIM ($p=0.104$), which leads on the text datasets. Against the best baseline of each configuration, a reference that no single method attains, its advantage is 1.18 points with a 95\% interval of $[{-}0.18,\,2.48]$. The two modalities behave differently. On the vision pairs KF-T leads that reference by 1.87 points $[0.58,\,3.23]$, which is significant over datasets ($p=0.037$). On the text pairs TIM is ahead of KF-T by 1.6 points: its objective rewards predictions that are balanced across classes, and the text test sets are balanced or nearly so. Section~\ref{sec:analysis} traces the difference to how much within-class variation the classes of a task share.

The fermented geometry is not tied to our product. Table~\ref{tab:plugin} runs five of the baselines on $\psi^{(T)}$ instead of the encoder features, with unchanged hyperparameters. All five improve on average, by 2.7 to 6.9 points. The gains are largest on the vision pairs, between 3.4 and 7.6 points, and smaller on text, between 0.0 and 4.0.

\begin{figure}[t]
\centering
\includegraphics[width=\columnwidth]{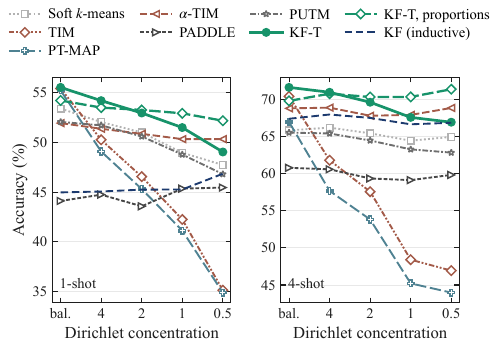}
\caption{Accuracy under class-imbalanced unlabeled sets.}
\label{fig:imbalance}
\end{figure}

\begin{table}[t]
\centering
\caption{Accuracy (\%) with $M$ unlabeled examples in total, with half of the classes absent, and with streamed unlabeled examples.}
\label{tab:protocol}
\footnotesize
\begin{tabular*}{\linewidth}{@{}l@{\extracolsep{\fill}}cccc@{}}
\toprule
Method & $M{=}20$ & $M{=}100$ & Half & Stream \\
\midrule
KF (inductive) & 57.8 & 56.4 & 57.0 & 56.2 \\
Soft $k$-means & 56.6 & 55.5 & 56.7 & 58.0 \\
BD-CSPN & 55.4 & 54.4 & 54.7 & 56.9 \\
LaplacianShot & 55.5 & 55.2 & 57.2 & 56.5 \\
TIM & 24.1 & 45.1 & 40.1 & 59.5 \\
PT-MAP & 42.6 & 44.5 & 38.7 & 57.4 \\
$\alpha$-TIM & 56.7 & 55.9 & 60.1 & 58.9 \\
PADDLE & 54.1 & 55.8 & 54.6 & 53.9 \\
\midrule
KF-T & \textbf{58.8} & \textbf{58.2} & 58.1 & \textbf{61.6} \\
KF-T, proportions & 56.8 & 56.3 & \textbf{61.9} & 60.0 \\
\bottomrule
\end{tabular*}
\end{table}

Table~\ref{tab:encoder} and Figure~\ref{fig:heatmap} break the results down by encoder and by pair, and \ref{app:results} reports them by dataset. KF is the best or tied best inductive method for every one of the 8 encoders. KF-T is the most accurate method for each of the five vision encoders, including the self-supervised DINOv2 on which the inductive geometry had little to change: once thousands of carriers are available, even mild anisotropy is estimated reliably and becomes worth removing. For the three sentence encoders TIM is the most accurate method. Among the datasets, Flowers-102 stands out because its test set is far from balanced: TIM and PT-MAP fall some twenty points or more below the inductive methods there, while KF-T and $\alpha$-TIM gain.

The 29 pairs that played no part in development give the same picture. On them KF-T leads the best baseline of each configuration by 1.75 points $[{-}0.07,\,3.68]$, against 0.68 points on the development pairs, and KF differs from the best inductive baseline by 0.12 and 0.03 points respectively (Table~\ref{tab:heldout}). The results are therefore not a product of choices made on the evaluation data.

Balanced unlabeled sets are a property of benchmarks rather than of applications \cite{veilleux2022realistic}. Figure~\ref{fig:imbalance} repeats the comparison on 8 pairs with unlabeled sets of 20 examples per class on average whose class proportions are drawn from a Dirichlet distribution; a smaller concentration means a stronger imbalance. For $\alpha$-TIM, which was designed for this setting, and PADDLE, which was designed for query sets that contain only a few of the classes, the better of the tuned and the authors' default configuration is reported at each concentration, and PUTM \cite{tian2023putm} is run with its authors' configuration for imbalanced queries. Methods that favor balanced predictions degrade quickly: between the balanced setting and concentration $0.5$ TIM loses 21.9 points and PT-MAP 21.7. KF-T does not constrain how its predictions are distributed and loses 5.6 points. It is the most accurate method down to concentration 1, leading $\alpha$-TIM by 2.0 points at concentration 2, the setting of \cite{veilleux2022realistic}, and by 1.1 at concentration 1; at concentration $0.5$, $\alpha$-TIM is ahead by 1.6 points. The cause is that the soft assignments \eqref{eq:post} treat all classes as equally likely. When the class proportions are estimated within the refinement passes, as the mean assignment of the unlabeled examples, the loss between the balanced setting and concentration $0.5$ shrinks to 0.2 points, and this variant is the most accurate method at concentrations 2, 1 and $0.5$, with 61.8\% against 59.6\% at the last. It costs 1.6 points in the balanced setting, which is why the uniform assignment remains the default.

Table~\ref{tab:protocol} varies the composition of the unlabeled set further on the same pairs, with averages over one and four shots and again the better configuration of $\alpha$-TIM and PADDLE in every column. With only 20 or 100 unlabeled examples for the whole task, drawn at random so that many classes are absent, KF-T is the most accurate method and stays above its inductive form, with 58.8\% against 57.8\% and 58.2\% against 56.4\%; TIM and PT-MAP, which spread their predictions over all classes, drop to 24.1\% and 42.6\% with 20 examples. When half of the classes are missing from an unlabeled set of 20 examples per present class, the margin of KF-T over its inductive form shrinks to about one point and $\alpha$-TIM is ahead, with 60.1\% against 58.1\%; the variant with estimated proportions is the most accurate method with 61.9\%. Estimating proportions requires enough unlabeled examples, however: with 20 or 100 in total that variant falls below the default. In the last column the unlabeled examples arrive in five chunks and the predictions for a chunk are final on arrival. KF-T attains 61.6\% under this constraint, against 63.6\% when the whole stream is available at once, and remains the most accurate method.

\subsection{Analysis}
\label{sec:analysis}

\begin{table}[t]
\centering
\caption{Components and spectral estimators (\%) on 11 pairs.}
\label{tab:ablation}
\footnotesize
\begin{tabular*}{\linewidth}{@{}l@{\extracolsep{\fill}}cccccc@{}}
\toprule
Inductive &  & 2 & 4 & 8 & 16 & Mean \\
\midrule
KF &  & 58.7 & 67.6 & 73.8 & 78.2 & 69.6 \\
w/o geometry &  & 58.1 & 66.5 & 72.8 & 77.1 & 68.6 \\
w/o refinement &  & 58.5 & 67.2 & 73.3 & 77.6 & 69.2 \\
w/o gate &  & 58.8 & 67.1 & 73.4 & 78.0 & 69.3 \\
in-sample variances &  & 58.4 & 67.3 & 73.5 & 78.1 & 69.4 \\
in-sample, own level &  & 58.9 & 67.8 & 74.0 & 78.3 & 69.7 \\
shrinkage variances &  & 58.4 & 67.3 & 73.5 & 78.0 & 69.3 \\
fold-wise directions &  & 58.9 & 67.5 & 73.8 & 78.1 & 69.6 \\
oracle variances &  & 58.7 & 67.6 & 73.7 & 78.1 & 69.5 \\
oracle covariance &  & 63.2 & 71.2 & 76.6 & 80.0 & 72.7 \\
\midrule
Transductive & 1 & 2 & 4 & 8 & 16 & Mean \\
\midrule
KF-T & 59.7 & 69.8 & 74.4 & 76.8 & 79.0 & 72.0 \\
w/o geometry & 56.1 & 64.8 & 69.2 & 72.1 & 74.6 & 67.4 \\
assignment evidence only & 59.8 & 69.9 & 74.4 & 76.6 & 78.7 & 71.9 \\
labeled evidence only & 50.8 & 62.2 & 69.9 & 75.0 & 78.6 & 67.3 \\
cold start & 59.7 & 69.3 & 73.6 & 75.9 & 78.3 & 71.4 \\
in-sample variances & 59.6 & 69.6 & 74.2 & 76.5 & 78.8 & 71.7 \\
shrinkage variances & 59.5 & 69.5 & 74.1 & 76.5 & 78.8 & 71.7 \\
fold-wise directions & 60.1 & 69.9 & 74.5 & 77.0 & 79.2 & 72.1 \\
true assignments & 74.0 & 80.1 & 82.6 & 83.8 & 84.7 & 81.0 \\
\bottomrule
\end{tabular*}
\end{table}

\begin{figure*}[t]
\centering
\includegraphics[width=\textwidth]{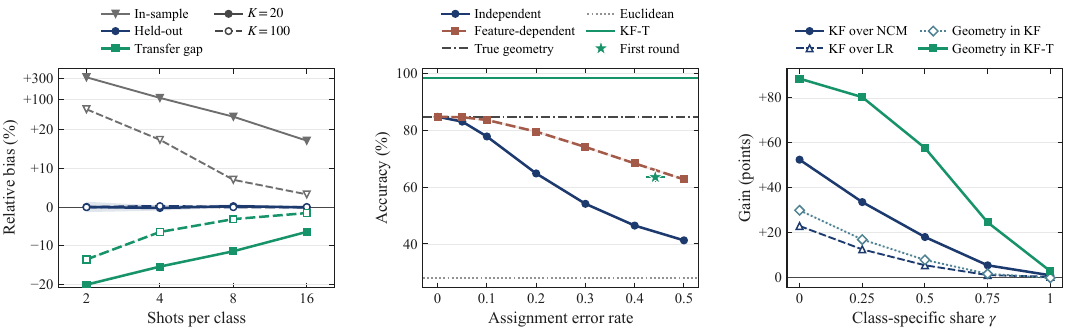}
\caption{Synthetic studies: bias of spectral estimates (left), accuracy under wrong assignments (middle), gains under class-specific covariances (right).}
\label{fig:synthetic}
\end{figure*}

\begin{figure}[t]
\centering
\includegraphics[width=\columnwidth]{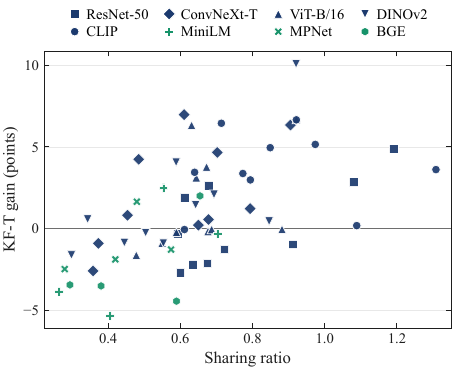}
\caption{Gain of KF-T over the best transductive baseline versus the sharing of within-class variation.}
\label{fig:structure}
\end{figure}

\begin{figure}[t]
\centering
\includegraphics[width=0.92\columnwidth]{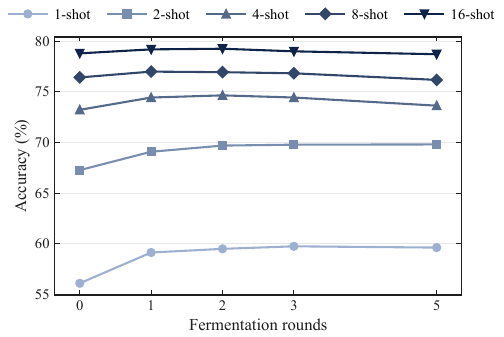}
\caption{Accuracy over fermentation rounds.}
\label{fig:rounds}
\end{figure}

\begin{figure}[t]
\centering
\includegraphics[width=\columnwidth]{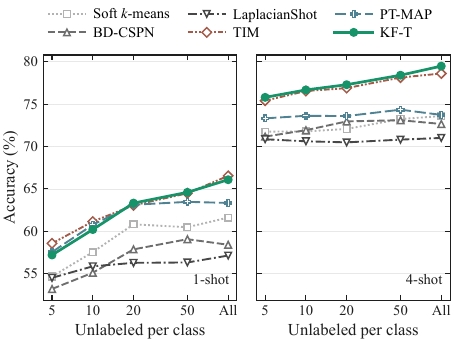}
\caption{Accuracy versus unlabeled examples per class.}
\label{fig:unlabeled}
\end{figure}

\begin{figure*}[t]
\centering
\includegraphics[width=0.96\textwidth]{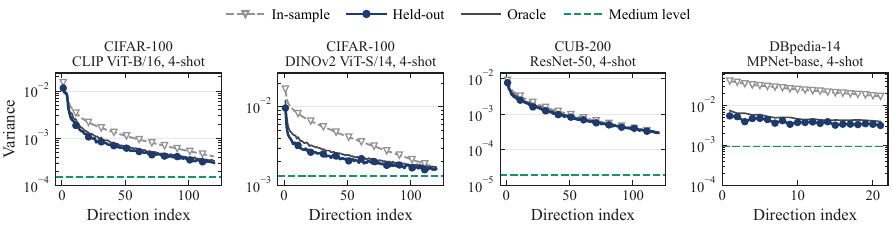}
\caption{Variance along estimated nuisance directions.}
\label{fig:spectrum}
\end{figure*}

\begin{figure*}[t]
\centering
\raisebox{1.7pt}{\includegraphics[scale=1.04]{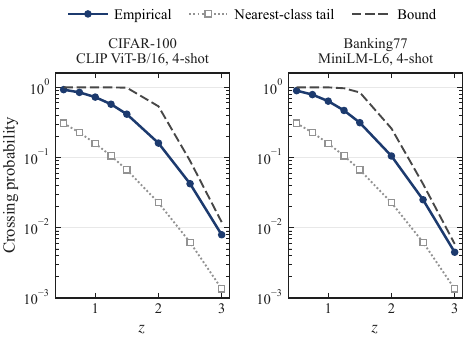}}\hspace{0.05\textwidth}\includegraphics[scale=1.04]{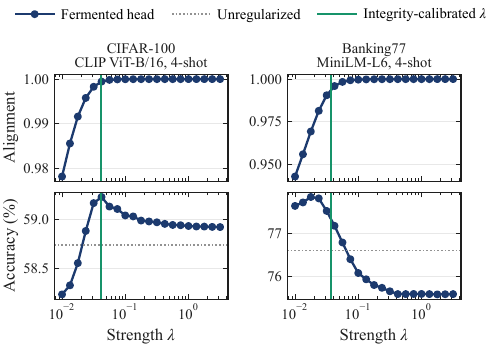}
\caption{Probability that a fermented strain changes class (left) and the fermented risk along the strength path (right).}
\label{fig:product}
\end{figure*}

Table~\ref{tab:ablation} removes or replaces one component at a time. Without the fermented geometry KF loses 1.0 points on average, without refinement by the fermented risk 0.4, and without the integrity gate 0.3. The lower rows of each block hold everything fixed except the spectral estimate: the directions, the rule for the medium level, the gate and the classifier are those of KF. Replacing the held-out variances by in-sample eigenvalues or by the eigenvalues of a Ledoit-Wolf shrinkage estimate changes the average accuracy of KF by 0.2 and 0.3 points, in-sample eigenvalues with their own mean as the level give 69.7\%, and the true variances of the same directions, measured on the full training pool, change the average by 0.1. With labeled carriers alone the map thus depends far more on which directions are damped than on the variances assigned to them, and what would help is better directions: the within-class covariance of the full training pool raises the average by 3.1 points. Averaging fold-wise spectral estimates, for which the unbiasedness of Proposition~\ref{prop:spectrum} holds without any transfer, gives 69.6\% against 69.6\%, so the transfer step costs nothing in accuracy. The held-out measurement therefore earns its place by calibration and not by accuracy: it is unbiased, it supplies the medium level that the in-sample variant borrows here, and it damps nothing when no structure exists, as the synthetic study below shows.

For KF-T the geometry is the dominant factor. Rounds that only update prototypes and assignments reach 67.4\% on average instead of 72.0\%. The spectral estimate again matters little, with 71.7\% and 71.7\% for in-sample and shrinkage variances and 72.1\% for fold-wise directions, whereas carriers formed with the true classes of the unlabeled examples would give 81.0\%: the accuracy of the assignments, not the estimation of the spectrum, limits the transductive geometry. A cold start from the Euclidean geometry costs 0.6 points, so the rounds depend only weakly on the labeled warm start. The assignments provide most of the gain at low shot counts; at sixteen shots the two sources of evidence in \eqref{eq:fusion} are about equally accurate and their fusion is ahead of both.

Figure~\ref{fig:synthetic} repeats three questions on Gaussian data whose covariance is known. The left panel concerns the spectrum of a covariance with eight strong shared directions in 256 dimensions (500 replications per setting). In-sample eigenvalues overstate the variance of their own directions by 3\% to 324\%, depending on the number of classes and shots. The held-out values deviate from the variance of the fold directions by at most 1.3\% in any setting, which is within sampling error, as \eqref{eq:unbiased} requires. Carried over to the directions of the full scatter they fall short by 1.5\% to 20.2\% along the strong directions, with a negative sign in 78\% to 96\% of the cases, so the transfer makes the damping cautious rather than aggressive; the worst-case bound \eqref{eq:transfer} held in all 675{,}000 checks and is far from tight. In accuracy the variance estimates are close to one another. With twenty classes and four shots the held-out, in-sample and shrinkage variances give 67.0\%, 68.6\% and 68.9\%, with a hundred classes 62.7\%, 60.0\% and 58.2\%, whereas the true covariance gives 84.2\% in the first case: what limits the geometry is the estimation of the directions, not of their variances. The estimates differ where no structure exists. For an isotropic covariance the held-out spectrum leaves the geometry essentially untouched and costs at most 0.08 points, while in-sample eigenvalues and linear shrinkage damp directions that carry no excess variance and lose up to 4.4 and 4.5 points.

The middle panel forms the carriers of 2,000 unlabeled points with a controlled fraction of wrong assignments and measures nearest-prototype accuracy in the resulting geometry (200 replications). With correct assignments the accuracy is 84.7\%, against 28.1\% in the Euclidean geometry. Wrong assignments drawn independently of the features reduce it to 64.8\% at an error rate of 0.2 and to 41.3\% at 0.5. Errors confined to the points nearest to a class boundary, the kind that pseudo-labels make, reduce it only to 79.5\% and 62.7\%, and the assignments of the first round of KF-T, wrong for 44\% of the points, yield 63.5\%. In every setting the second moment of the carriers stayed below the bound \eqref{eq:leak} evaluated with the realized noise level, and the accuracy stayed above the Euclidean one. The damping thus limits itself as Proposition~\ref{prop:rounds} states, although it does not vanish. The complete KF-T reaches 98.2\% on these tasks, 99.0\% when the true assignments are supplied, and 16.6\% without the geometry.

The right panel mixes the shared covariance with a class-specific one of the same spectrum, whose strong directions are drawn independently for every class. All gains decrease monotonically with the class-specific share $\gamma$. For twenty classes and four shots the gain of KF over NCM falls from 52.6 points at $\gamma=0$ to 1.1 at $\gamma=1$, and the contribution of the geometry to KF-T from 88.6 to 3.0 points. At $\gamma=1$ the geometry changes the accuracy of KF by less than $0.1$ points, and the gate lowers its average level from 1.00 to 0.82. When the assumption of a shared covariance fails, the method loses its advantage but does not turn harmful.

Model \eqref{eq:model} assumes a covariance shared by all classes, and the benchmark allows a direct test. Class by class, the ten leading directions of the within-class covariance pooled over all other classes capture on average 23\% of the held-out variance of the class, against 36\% for the ten leading directions of the class itself and 1.5\% for random directions (\ref{app:results}). Exact sharing is thus rejected in every pair, but a common low-dimensional structure exists, and it is more pronounced for the vision encoders than for the sentence encoders: the ratio of the first two numbers, the sharing ratio, averages 0.70 and 0.47. This ratio predicts where fermentation pays (Figure~\ref{fig:structure}). Across the 62 pairs its rank correlation with the gain of KF-T over the best transductive baseline is 0.60, and 0.52 when the difficulty of the task is controlled for. KF-T trails the best baseline in 129 of the 310 configurations, by 2.0 points on average. These include 41 of the 60 text configurations and 88 of the 250 vision configurations, and the pairs concerned have a lower sharing ratio than the others, with a median of 0.56 against 0.69. The limits of the method are those of its premise: where classes do not vary along common directions there is little second-order knowledge to exchange, and a rule that balances its predictions is the better choice.

Figure~\ref{fig:rounds} follows accuracy over the rounds. With the warm-start geometry alone KF-T reaches 70.4\% on average; one round gives 71.8\%, two 72.0\%, three 72.0\% and five 71.6\%. Most of the benefit thus arrives in the first round and the result is stable afterwards. Figure~\ref{fig:unlabeled} varies the size of the unlabeled set. With one shot, KF-T improves on the inductive form by 5.6 points with five unlabeled examples per class and by 14.3 points with fifty.

Figure~\ref{fig:spectrum} tests Proposition~\ref{prop:spectrum} on the benchmark. For four pairs at four shots it shows, for the leading estimated directions, the in-sample eigenvalue, the held-out variance \eqref{eq:heldout}, and the variance measured with all training examples of the dataset, which serves as an oracle. Over the first twenty directions the in-sample spectrum overstates the oracle by factors between 1.17 and 5.20, whereas the held-out spectrum stays between 0.79 and 0.91 of it. Evaluating the oracle on the fold directions as well separates the two sources of the remaining gap. Relative to that fold-level target the held-out values average 0.89 to 0.97, where model \eqref{eq:model} predicts one: directions learned on one group of classes carry slightly less variance on the other group, which measures how far the class covariances are from being shared. The fold-level target in turn amounts to 0.88 to 0.93 of the oracle along the directions of the full scatter, the cost of the transfer in \eqref{eq:transferbias}. Both effects make the matured spectrum conservative. The figure also shows why encoders respond differently: with CLIP features the leading directions exceed the medium level by almost two orders of magnitude, so damping them changes the geometry substantially.

Figure~\ref{fig:product} examines the product. The measured probability that a noisy strain becomes closer to another strain lies below the bound \eqref{eq:union} for every integrity level. Across all 62 pairs at four shots and $z=2$, a fermented strain ends closer to another strain with probability 10.9\% on average, 7.2\% for tasks with at most 20 classes and 16.1\% for tasks with at least 100, where the largest value is 22.7\%. The pairwise level of $2.3\%$ thus does not carry over to the class as a whole when many strains lie near the nearest distance. The level $z=\bar{\Phi}^{-1}(0.05/(K-1))$, between 2.45 and 3.48 on the benchmark, lowers the measured probability to 0.8\% on average and 1.9\% at most. On the 6 pairs of the sensitivity study this level changes the accuracy of KF by $-0.2$ points and that of KF-T by 0.2, so a guarantee for the class as a whole is available at no measurable cost in accuracy. Along the strength path the weights align with the centered prototypes as Theorem~\ref{thm:risk} states; the cosine similarity between the two reaches 1.00 for strong fermentation. Accuracy peaks at intermediate strengths, and the integrity-calibrated strength falls in that region: it attains 59.2\% and 77.2\% on the two pairs, against best values of 59.2\% and 77.9\% over the path and 58.7\% and 76.6\% without regularization.

The remaining constants have little influence. Three or five class groups instead of two give 78.0\% and 78.0\% for KF, against 78.0\%, and 76.5\% and 76.4\% for KF-T, against 76.4\%. For KF the integrity level is equally uncritical, with accuracies between 77.5\% and 78.0\% for $z$ from 1 to 4. For KF-T the level also sets the width of the assignment kernel \eqref{eq:post}: accuracy is 76.4\% at $z=2$, 76.8\% at $z=3$ and 76.7\% at $z=4$, and decreases to 74.6\% at $z=1.5$ and 69.7\% at $z=1$, where the kernel is too broad to separate neighboring classes. A finer grid of five gate levels leaves the average accuracy unchanged to within $0.1$ points, and five refinement passes instead of three change it by 0.1; a single pass loses 2.1 points, most of them at one shot, where the prototypes have then barely moved from the labeled strains. The computational cost is small. Averaged over all configurations KF takes 0.19~s per episode and KF-T 0.96~s on a single GPU, compared with 0.09~s for logistic regression and 0.91~s for TIM.

\section{Conclusion}
\label{sec:conclusion}

Low-data classification on frozen features is limited by missing knowledge about within-class variation, and that knowledge can be produced inside the task. Knowledge Fermentation pools the residuals of all classes, keeps the directions whose variance is confirmed on classes that did not define them, converts them into a change of geometry that is equivalent, for a linear classifier, to training under the structured noise, and trains a convex classifier at a noise strength calibrated on the distances between prototypes. Unlabeled examples enter the same exchange through fermentation rounds. Each step is backed by a statement of what it guarantees: unbiasedness of the held-out variances with a bounded transfer, one-sidedness of the fermented map, a pairwise bound on identity violations whose total can be controlled, the two ends of the objective between the linear probe and a prototype rule, and a floor on the separation that survives wrong assignments. Across 14 datasets and 8 encoders the method runs with fixed constants and is on par with the best tuned inductive baselines. With unlabeled carriers it lifts one-shot accuracy by 13.4 points and is the most accurate transductive method on average, significantly so on vision encoders, whereas on sentence encoders, whose classes share little of their variation, information maximization remains ahead. The fermented geometry is a component in its own right, since it raises the accuracy of other transductive rules that are run inside it. The same principle applies wherever classes share nuisance structure and labels are the scarce resource.

\section*{Code availability}

The code, checkpoints and artifacts of this study will be released upon publication.

\appendix
\setcounter{table}{0}
\setcounter{figure}{0}
\renewcommand{\theHtable}{app\Alph{section}.\arabic{table}}
\renewcommand{\theHfigure}{app\Alph{section}.\arabic{figure}}

\section{Additional results}
\label{app:results}

\begin{figure*}[t]
\centering
\includegraphics[width=0.9\textwidth]{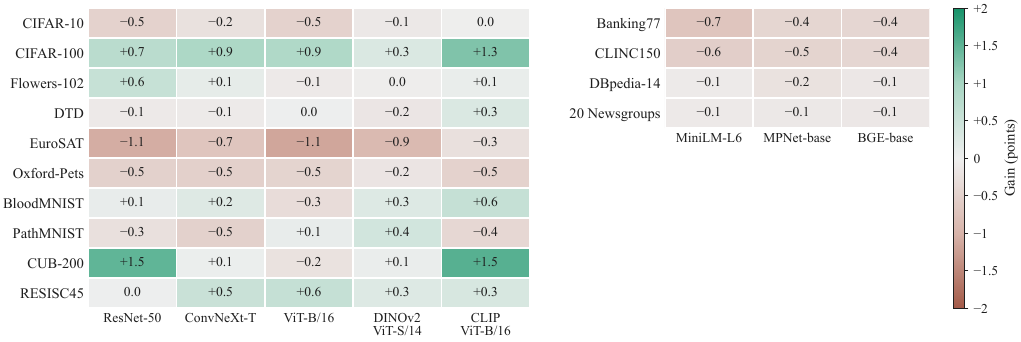}
\caption{Gain of KF over the best inductive baseline (points), averaged over shots.}
\label{fig:indheat}
\end{figure*}

Table~\ref{tab:dataset} reports accuracy by dataset and Figure~\ref{fig:indheat} the gain of KF over the best inductive baseline for every pair. Table~\ref{tab:statistics} lists the paired comparisons with their intervals and tests, Table~\ref{tab:heldout} separates development from held-out pairs, and Table~\ref{tab:covariance} describes the within-class covariance for every encoder.

\begin{table*}[t]
\centering
\caption{Accuracy by dataset (\%), averaged over encoders and shots.}
\label{tab:dataset}
\small
\begin{tabular}{@{}lccccccc@{}}
\toprule
Method & CIFAR-10 & CIFAR-100 & Flowers & DTD & EuroSAT & Pets & Blood \\
\midrule
NCM & 79.3 & 55.5 & 87.9 & 52.6 & 66.9 & 80.8 & 54.1 \\
LR & \underline{80.0} & 57.8 & 89.7 & 54.8 & 71.5 & 81.7 & 58.7 \\
LDA & 79.6 & 58.1 & 90.0 & 53.9 & 71.3 & \textbf{82.6} & 59.4 \\
Noise & 79.7 & 57.0 & 89.1 & 54.0 & 70.8 & 80.7 & 58.4 \\
Mixup & \underline{80.0} & 57.4 & 89.3 & 54.4 & \underline{71.6} & 80.9 & 59.2 \\
ISDA & \textbf{80.1} & \underline{58.5} & \underline{90.2} & \underline{55.1} & \textbf{72.0} & 82.1 & \underline{59.7} \\
KF & 79.9 & \textbf{59.6} & \textbf{90.5} & \textbf{55.2} & 71.2 & \underline{82.3} & \textbf{60.0} \\
\midrule
Soft $k$-means & 85.9 & 57.7 & 90.4 & 53.8 & 67.7 & 85.9 & 51.5 \\
BD-CSPN & 81.8 & 58.1 & 89.5 & 54.4 & 69.5 & 85.3 & 52.7 \\
LaplacianShot & 85.0 & 56.6 & 89.0 & 54.4 & 70.2 & 83.3 & 54.4 \\
TIM & \textbf{90.2} & \underline{63.8} & 70.6 & \underline{58.2} & 77.3 & \textbf{90.0} & 54.8 \\
PT-MAP & 87.2 & 58.1 & 65.0 & 54.1 & 65.0 & 87.0 & 47.4 \\
PT-MAP (power) & 87.8 & 60.1 & 65.4 & 55.4 & 66.8 & 88.5 & 47.7 \\
$\alpha$-TIM & 88.9 & 62.4 & \underline{93.4} & 57.3 & \underline{78.9} & 87.8 & \underline{61.4} \\
PADDLE & 78.7 & 48.9 & 85.3 & 49.5 & 59.8 & 79.8 & 44.3 \\
protoLP & 88.0 & 58.4 & 62.8 & 53.8 & 63.0 & 88.4 & 45.0 \\
KF-T & \underline{89.2} & \textbf{67.3} & \textbf{94.1} & \textbf{58.8} & \textbf{79.6} & \underline{89.6} & \textbf{67.2} \\
\midrule
Method & Path & CUB & RESISC & Banking & CLINC & DBpedia & 20NG \\
\midrule
NCM & 63.3 & 64.0 & 57.8 & 74.6 & 84.2 & 75.5 & \textbf{51.0} \\
LR & \underline{66.5} & 66.0 & 61.5 & \underline{76.4} & \underline{85.1} & 75.9 & \underline{50.5} \\
LDA & 66.3 & 65.9 & 61.3 & 74.9 & 84.0 & 76.1 & 49.4 \\
Noise & 65.6 & 66.0 & 60.6 & 76.1 & 84.9 & 75.2 & 49.6 \\
Mixup & 66.4 & 65.7 & 61.1 & 75.8 & 84.7 & 75.6 & 50.1 \\
ISDA & \textbf{66.8} & \underline{66.7} & \underline{62.4} & \textbf{76.7} & \textbf{85.3} & \underline{76.2} & 50.1 \\
KF & \textbf{66.8} & \textbf{67.7} & \textbf{62.8} & 76.2 & 84.8 & \textbf{76.3} & \textbf{51.0} \\
\midrule
Soft $k$-means & 65.9 & 67.6 & 60.5 & 77.5 & 89.4 & 89.9 & 57.6 \\
BD-CSPN & 65.3 & 67.5 & 60.4 & 77.1 & 88.1 & 80.4 & 55.7 \\
LaplacianShot & 66.2 & 63.5 & 60.0 & 76.6 & 86.0 & 83.6 & 55.3 \\
TIM & 60.9 & \underline{71.5} & \underline{68.7} & \textbf{82.5} & \textbf{92.3} & \textbf{94.6} & \underline{59.5} \\
PT-MAP & 54.6 & 67.4 & 60.1 & 79.1 & \underline{91.8} & 91.1 & 58.4 \\
PT-MAP (power) & 54.7 & 69.0 & 61.1 & 78.8 & 91.5 & 90.8 & 58.4 \\
$\alpha$-TIM & \underline{71.0} & 70.2 & 68.1 & \underline{80.2} & 89.5 & 90.7 & 58.1 \\
PADDLE & 57.7 & 58.7 & 45.3 & 70.8 & 80.5 & 75.3 & 43.8 \\
protoLP & 52.7 & 68.3 & 59.5 & 77.7 & 91.2 & 91.5 & 57.0 \\
KF-T & \textbf{74.0} & \textbf{73.9} & \textbf{73.4} & 78.9 & 89.2 & \underline{92.6} & \textbf{61.7} \\
\bottomrule
\end{tabular}
\end{table*}

\begin{table*}[t]
\centering
\caption{Paired differences in accuracy (points): bootstrap interval over datasets, effect size $d$, and $p$-values over configurations and datasets.}
\label{tab:statistics}
\small
\begin{tabular}{@{}lcccccc@{}}
\toprule
Comparison & $\Delta$ & 95\% CI & $d$ & W/T/L & $p_\mathrm{cfg}$ & $p_\mathrm{ds}$ \\
\midrule
\multicolumn{7}{@{}l}{\emph{Inductive, 2--16 shots}} \\
KF vs.\ NCM & $+3.56$ & $[+2.38, +4.65]$ & $0.97$ & 225/3/20 & $<10^{-10}$ & $2.4{\times}10^{-4}$ \\
KF vs.\ LR & $+0.85$ & $[+0.38, +1.32]$ & $0.59$ & 169/21/58 & $<10^{-10}$ & 0.005 \\
KF vs.\ LDA & $+0.98$ & $[+0.54, +1.42]$ & $0.79$ & 191/9/48 & $<10^{-10}$ & 0.001 \\
KF vs.\ Noise & $+1.34$ & $[+0.85, +1.79]$ & $0.85$ & 206/7/35 & $<10^{-10}$ & $8.5{\times}10^{-4}$ \\
KF vs.\ Mixup & $+0.99$ & $[+0.49, +1.47]$ & $0.68$ & 186/11/51 & $<10^{-10}$ & 0.003 \\
KF vs.\ ISDA & $+0.32$ & $[-0.06, +0.69]$ & $0.30$ & 131/16/101 & $5.7{\times}10^{-4}$ & 0.217 \\
KF vs.\ best inductive & $+0.07$ & $[-0.26, +0.40]$ & $0.08$ & 110/17/121 & 0.876 & 0.952 \\
\midrule
\multicolumn{7}{@{}l}{\emph{Transductive, 1--16 shots}} \\
KF-T vs.\ Soft $k$-means & $+6.82$ & $[+4.43, +9.19]$ & $1.12$ & 287/1/22 & $<10^{-10}$ & $2.4{\times}10^{-4}$ \\
KF-T vs.\ BD-CSPN & $+7.65$ & $[+5.67, +9.60]$ & $1.36$ & 304/2/4 & $<10^{-10}$ & $1.2{\times}10^{-4}$ \\
KF-T vs.\ LaplacianShot & $+7.81$ & $[+5.99, +9.56]$ & $1.39$ & 303/1/6 & $<10^{-10}$ & $1.2{\times}10^{-4}$ \\
KF-T vs.\ TIM & $+4.61$ & $[+0.89, +8.90]$ & $0.58$ & 204/5/101 & $<10^{-10}$ & 0.104 \\
KF-T vs.\ PT-MAP & $+9.87$ & $[+5.24, +14.58]$ & $0.99$ & 272/1/37 & $<10^{-10}$ & 0.001 \\
KF-T vs.\ PT-MAP (power) & $+9.06$ & $[+4.43, +13.95]$ & $0.90$ & 271/1/38 & $<10^{-10}$ & 0.001 \\
KF-T vs.\ $\alpha$-TIM & $+2.41$ & $[+1.31, +3.50]$ & $0.70$ & 224/7/79 & $<10^{-10}$ & 0.002 \\
KF-T vs.\ PADDLE & $+15.33$ & $[+12.17, +18.64]$ & $1.54$ & 306/0/4 & $<10^{-10}$ & $1.2{\times}10^{-4}$ \\
KF-T vs.\ protoLP & $+10.47$ & $[+5.32, +15.83]$ & $0.95$ & 265/2/43 & $<10^{-10}$ & 0.001 \\
KF-T vs.\ best transductive & $+1.18$ & $[-0.18, +2.48]$ & $0.33$ & 176/7/127 & $2.3{\times}10^{-6}$ & 0.296 \\
\midrule
KF-T vs.\ KF & $+7.38$ & $[+5.92, +8.91]$ & $1.15$ & 291/7/12 & $<10^{-10}$ & $1.2{\times}10^{-4}$ \\
\bottomrule
\end{tabular}
\end{table*}

\begin{table*}[t]
\centering
\caption{Differences to the best baseline of each configuration (points) on development and held-out pairs.}
\label{tab:heldout}
\small
\begin{tabular}{@{}llccccc@{}}
\toprule
 & & \multicolumn{2}{c}{Development} & \multicolumn{2}{c}{Held-out} & Held-out $-$ dev. \\
\cmidrule(lr){3-4}\cmidrule(lr){5-6}\cmidrule(lr){7-7}
Pairs & Comparison & $\Delta$ [95\% CI] & $n$ & $\Delta$ [95\% CI] & $n$ & $\Delta$ [95\% CI] \\
\midrule
All & KF vs.\ best inductive (2--16) & $+0.03$ $[-0.41, +0.57]$ & 33 & $+0.12$ $[-0.11, +0.38]$ & 29 & $+0.10$ $[-0.45, +0.56]$ \\
 & KF-T vs.\ best transductive (1--16) & $+0.68$ $[-0.74, +2.09]$ & 33 & $+1.75$ $[-0.07, +3.68]$ & 29 & $+1.07$ $[-0.88, +3.34]$ \\
\midrule
Vision & KF vs.\ best inductive (2--16) & $+0.24$ $[-0.40, +1.08]$ & 22 & $+0.13$ $[-0.10, +0.40]$ & 28 & $-0.10$ $[-0.97, +0.53]$ \\
 & KF-T vs.\ best transductive (1--16) & $+1.74$ $[+0.68, +3.33]$ & 22 & $+1.97$ $[+0.18, +4.08]$ & 28 & $+0.23$ $[-1.84, +2.26]$ \\
\bottomrule
\end{tabular}
\end{table*}

\begin{table*}[t]
\centering
\caption{Within-class covariance by encoder, with values under exact sharing (Ref.), and average gains (points).}
\label{tab:covariance}
\small
\begin{tabular}{@{}lcccccccc@{}}
\toprule
 & & & \multicolumn{2}{c}{Sharing ratio, $k=10$} & \multicolumn{2}{c}{Heterogeneity} & \multicolumn{2}{c}{Gain (points)} \\
\cmidrule(lr){4-5}\cmidrule(lr){6-7}\cmidrule(lr){8-9}
Encoder & Pairs & Anisotropy & Obs. & Ref. & Obs. & Ref. & KF $-$ LR & KF-T $-$ best \\
\midrule
ResNet-50 & 10 & 2518.0 & 0.77 & 1.53 & 0.42 & 0.21 & $+0.92$ & $+0.27$ \\
ConvNeXt-T & 10 & 119.1 & 0.60 & 1.72 & 0.46 & 0.21 & $+0.73$ & $+2.16$ \\
ViT-B/16 & 10 & 137.2 & 0.67 & 1.73 & 0.43 & 0.21 & $+0.85$ & $+1.70$ \\
DINOv2 ViT-S/14 & 10 & 54.2 & 0.58 & 1.54 & 0.48 & 0.21 & $+0.51$ & $+1.52$ \\
CLIP ViT-B/16 & 10 & 419.0 & 0.87 & 1.48 & 0.36 & 0.21 & $+2.12$ & $+3.68$ \\
\midrule
MiniLM-L6 & 4 & 19.4 & 0.48 & 1.55 & 0.42 & 0.16 & $+0.21$ & $-1.75$ \\
MPNet-base & 4 & 101.8 & 0.44 & 1.59 & 0.39 & 0.16 & $+0.20$ & $-0.99$ \\
BGE-base & 4 & 68.9 & 0.48 & 1.62 & 0.41 & 0.16 & $-0.02$ & $-2.33$ \\
\midrule
All pairs & 62 & 182.8 & 0.65 & 1.60 & 0.42 & 0.20 & $+0.85$ & $+1.18$ \\
\bottomrule
\end{tabular}
\end{table*}

\section{Implementation details}
\label{app:details}
\setcounter{table}{0}
\setcounter{figure}{0}

This appendix records what is needed to reproduce the experiments. Images are resized to 256 pixels on the shorter side with bicubic interpolation and center-cropped to 224; the feature is the pooled output of the encoder without its classification head, which for CLIP is the 768-dimensional output before the projection to the joint image-text space. The CLIP weights are run with the exact GELU activation of the model definition named in Table~\ref{tab:checkpoints}, whereas the original implementation uses the approximation $x\,\sigma(1.702\,x)$, and DINOv2, whose checkpoint is defined for inputs of 518 pixels, is run at 224 pixels with interpolated position embeddings. Text is truncated to 256 tokens and embedded with the pooling and the final $\ell_2$ normalization of the respective sentence encoder. Table~\ref{tab:checkpoints} lists the public checkpoints. The training pool of a dataset is its training split, joined with the validation split where one exists. EuroSAT has no standard split and is divided at random, stratified by class, into 80\% for the pool and 20\% for testing; RESISC45 is used with the split of its copy in the timm collection and DTD with its first split. Training pools and test sets are drawn once with a fixed seed. For every shot count the support sets are drawn with a fixed seed and shared by all methods, so every comparison is paired, and the class groups of KF are drawn from the same seed.

\begin{table}[t]
\centering
\caption{Encoder checkpoints.}
\label{tab:checkpoints}
\footnotesize
\begin{tabular*}{\linewidth}{@{}l@{\extracolsep{\fill}}lr@{}}
\toprule
Encoder & Checkpoint & $d$ \\
\midrule
ResNet-50 & \ckpt{resnet50.tv\_in1k} & 2048 \\
ConvNeXt-T & \ckpt{convnext\_tiny.fb\_in22k\_ft\_in1k} & 768 \\
ViT-B/16 & \ckpt{vit\_base\_patch16\_224.augreg\_in21k\_ft\_in1k} & 768 \\
DINOv2 ViT-S/14 & \ckpt{vit\_small\_patch14\_dinov2.lvd142m} & 384 \\
CLIP ViT-B/16 & \ckpt{vit\_base\_patch16\_clip\_224.openai} & 768 \\
MiniLM-L6 & \ckpt{all-MiniLM-L6-v2} & 384 \\
MPNet-base & \ckpt{all-mpnet-base-v2} & 768 \\
BGE-base & \ckpt{bge-base-en-v1.5} & 768 \\
\bottomrule
\end{tabular*}
\end{table}

The fermented risk \eqref{eq:risk} is minimized from zero weights with L-BFGS, using a history of 20 updates and a strong Wolfe line search, for at most 200 iterations in KF and 150 in KF-T; the search stops earlier when the largest gradient entry falls below $10^{-7}$ or the objective changes by less than $10^{-12}$. When $N<d$ the features are first expressed in an orthonormal basis of the span of the support set, which leaves the minimizer unchanged. Fermentation requires $n\ge2$ and at least $2F$ classes, and the gate at least $2F^2$ classes; every task of the benchmark satisfies both. Logistic regression and ISDA use the same optimizer, Noise and Mixup use 300 Adam steps with step size $0.05$ and cosine decay, and the transductive baselines use the iteration counts of their reference implementations unless the count is part of the grid in Table~\ref{tab:grids}. Image features are extracted with half-precision autocasting and stored in single precision; all other computations run in single precision on one NVIDIA RTX~5090 with PyTorch~2.11, and the reported times were measured on that device.

\begin{table}[t]
\centering
\caption{Hyperparameter grids of the baselines.}
\label{tab:grids}
\footnotesize
\begin{tabular*}{\linewidth}{@{}l@{\extracolsep{\fill}}p{0.7\linewidth}@{}}
\toprule
Method & Values \\
\midrule
LR & weight decay $10^{-6},\dots,10^{-2}$ \\
Noise & $\sigma\in\{0.1,0.3,1\}$; weight decay $10^{-5},10^{-4}$ \\
Mixup & $\beta\in\{0.2,1\}$; weight decay $10^{-5},10^{-4}$ \\
ISDA & strength $\{0.25,0.5,1,2\}$; weight decay $10^{-5},10^{-4},10^{-3}$ \\
Soft $k$-means & temperature $\{1,0.1,0.05,0.02\}$; steps $\{1,10\}$ \\
BD-CSPN & eight combinations of pseudo-labels per class $\{8,16,32\}$ and cosine scale $\varepsilon\in\{1,3,10\}$ \\
LaplacianShot & twelve combinations of neighbors $\{2,3,10\}$, $\lambda\in\{0.7,1,3,10,30\}$ and rectification on or off \\
TIM & eight combinations of temperature $\{7.5,15,30\}$, entropy weight $\{0.1,0.5,1\}$ and step size $10^{-4},10^{-3}$ \\
PT-MAP & $\lambda\in\{10,20,30,50\}$; $(\alpha,\text{steps})\in\{(0.4,30),(0.2,20)\}$ \\
$\alpha$-TIM & $\alpha\in\{1.1,1.25,1.5,2,3,4,5,7\}$ \\
PADDLE & $\lambda/M\in\{0.1,0.2,0.4,0.6,0.8,1,1.2\}$ \\
protoLP & transport weight $\{10,20,50\}$ with 8 retained dimensions per class; 16 per class at weight 20 \\
PUTM & authors' configuration for imbalanced queries \\
\bottomrule
\end{tabular*}
\end{table}

Every baseline was compared with reference code on identical inputs: five-way tasks with 75 queries for all methods and, except for protoLP, PUTM and the power-transform form of PT-MAP, also tasks with all classes drawn from pairs of the benchmark; the reference code of protoLP and PUTM is restricted to five classes. The ISDA loss and the transductive methods TIM, LaplacianShot, PT-MAP, $\alpha$-TIM, PADDLE, protoLP and PUTM were run against the code released by their authors, soft $k$-means against a port of the authors' code, BD-CSPN, whose authors released no code, against an independent implementation of the published equations, and LR and LDA against scikit-learn. Table~\ref{tab:validation} reports the share of queries on which the predictions coincide; it is at least 99.99\% for every method. Three conventions differ from the reference code without changing the algorithms: the neighbor count of LaplacianShot excludes the point itself, the step count of PT-MAP in its form without the power transform includes the final assignment, and the temperature of soft $k$-means amounts to rescaling the features. On the features released with PT-MAP \cite{hu2021leveraging} and the 10{,}000 tasks of its authors, our implementation returns the accuracies reported by the authors on miniImageNet (82.92\% and 88.82\% at one and five shots) and CIFAR-FS (87.69\% and 90.68\%), and on CUB those of the authors' code (91.63\% and 94.09\%, reported 91.55\% and 93.99\%).

\begin{table}[t]
\centering
\caption{Agreement of predictions with reference implementations (\%).}
\label{tab:validation}
\footnotesize
\begin{tabular*}{\linewidth}{@{}l@{\extracolsep{\fill}}lc@{}}
\toprule
Method & Reference & Agreement \\
\midrule
LR & scikit-learn & 99.99 \\
LDA & scikit-learn & 100.00 \\
ISDA & authors' code & 99.99 \\
Soft $k$-means & port of the authors' code & 100.00 \\
BD-CSPN & published equations & 100.00 \\
LaplacianShot & authors' code & 100.00 \\
LaplacianShot, rectified & authors' code & 100.00 \\
TIM & authors' code & 100.00 \\
PT-MAP & authors' code & 100.00 \\
PT-MAP, power transform & authors' code & 100.00 \\
$\alpha$-TIM & authors' code & 100.00 \\
PADDLE & authors' code & 100.00 \\
protoLP & authors' code & 99.998 \\
PUTM & authors' code & 100.00 \\
\bottomrule
\end{tabular*}
\end{table}

\section{Proofs}
\label{app:proofs}

The results of Section~\ref{sec:theory} are proved in order.

\begin{proof}[Proof of Proposition~\ref{prop:spectrum}]
Fix a class $c\in G_f$ and write its nuisance terms as $\xi_1,\dots,\xi_n$ and their mean as $\bar\xi$. Its carriers are $\xi_i-\bar\xi$, so $\sum_i(u^{\T}(\xi_i-\bar\xi))^2$ is the sum of squared deviations of the scalars $a_i=u^{\T}\xi_i$ from their mean. Conditionally on $R_{-f}$ the vector $u$ is fixed and the $a_i$ are independent with variance $u^{\T}\bS u$, since the carriers of $G_f$ are independent of those outside it. The expectation of the sum is therefore $(n-1)u^{\T}\bS u$. Summing over the $|G_f|$ classes and dividing by $\nu_f$ gives \eqref{eq:unbiased}. If $\xi$ is Gaussian, each sum is $u^{\T}\bS u$ times a $\chi^2_{n-1}$ variable, the classes are independent, and $\|R_fu\|^2/(u^{\T}\bS u)\sim\chi^2_{\nu_f}$, whose variance is $2\nu_f$.

For \eqref{eq:transfer}, the matrix $D=vv^{\T}-uu^{\T}$ is symmetric, has trace zero and vanishes on the orthogonal complement of the span of $u$ and $v$. On that span, in the basis $(v,u)$ and with $c=u^{\T}v$, it acts as the matrix with rows $(1,c)$ and $(-c,-1)$, whose eigenvalues are $\pm\sqrt{1-c^2}=\pm\sin\theta$. Hence $D=\sin\theta\,(aa^{\T}-bb^{\T})$ for orthonormal vectors $a$ and $b$, and $v^{\T}\bS v-u^{\T}\bS u=\tr(\bS D)=\sin\theta\,(a^{\T}\bS a-b^{\T}\bS b)$, which lies between $-(\ell_1-\ell_d)\sin\theta$ and $(\ell_1-\ell_d)\sin\theta$. Inequality \eqref{eq:transferbias} follows by applying this to $v_i$ and $v^{(-f)}_i$, taking expectations with \eqref{eq:unbiased}, and averaging over the groups.

For \eqref{eq:insample}, Ky Fan's maximum principle states that for a symmetric matrix $M$ the sum of its $k$ largest eigenvalues equals $\max\tr(V^{\T}MV)$ over $d\times k$ matrices $V$ with orthonormal columns. Let $V_\star$ hold the top $k$ eigenvectors of $\bS$. Then $\sum_{i\le k}\hat\ell_i\ge\tr(V_\star^{\T}SV_\star)$, and taking expectations with $\E S=\bS$ gives $\E\sum_{i\le k}\hat\ell_i\ge\sum_{i\le k}\ell_i$. Applying the principle to $\bS$ and the random frame $[v_1,\dots,v_k]$ gives $\sum_{i\le k}v_i^{\T}\bS v_i\le\sum_{i\le k}\ell_i$ for every realization. Finally, if $\nu<d$ then $S$ has rank at most $\nu$, so $\sum_{i\le\nu}\hat\ell_i=\tr S$ with expectation $\tr\bS=\sum_{i\le d}\ell_i$, and the difference to $\sum_{i\le\nu}\ell_i$ is $\sum_{i>\nu}\ell_i$.
\end{proof}

\begin{proof}[Proof of Proposition~\ref{prop:geometry}]
Let $V$ have the orthonormal columns $v_i$, $i\in I$. Then $\hbS/\hat m=(I_d-VV^{\T})+V\operatorname{diag}(h_i/\hat m)V^{\T}$, a spectral decomposition with eigenvalue $1$ on the complement of the span of $V$ and $h_i/\hat m=g_i^{-2}$ on $v_i$. Hence $(\hbS/\hat m)^{-s/2}=(I_d-VV^{\T})+V\operatorname{diag}(g_i^{\,s})V^{\T}=A_s$, which proves (a). For (b), $A_1\epsilon$ is Gaussian with covariance $(\lambda/\hat m)A_1\hbS A_1^{\T}=\lambda I_d$ by (a), and $\langle A_1^{-1}w,A_1(x+\epsilon)\rangle=\langle w,x+\epsilon\rangle$ because $A_1$ is symmetric and invertible; the logits of $(W,b)$ at $x+\epsilon$ therefore coincide with those of $(WA_1^{-1},b)$ at $A_1x+A_1\epsilon$. For (c), with equal priors the Bayes rule maximizes $-\tfrac12(x-\mu_c)^{\T}\hbS^{-1}(x-\mu_c)=-\tfrac{1}{2\hat m}\|A_1(x-\mu_c)\|^2$. For (d), $A_s$ is symmetric with eigenvalue $1$ on the complement of the span of $V$ and $g_i^{\,s}\in[g_{\min}^{\,s},1]$ on $v_i$, because $h_i>\hat m$ for $i\in I$. Its operator norm is at most one, its smallest eigenvalue is at least $g_{\min}^{\,s}$, and $g_i\le g_i^{\,s}\le1$ for $s\in[0,1]$ gives the ordering. For (e), $\langle A_sx,A_sx'\rangle=x^{\T}A_s^2x'=\hat m^{\,s}\,x^{\T}\hbS^{-s}x'$ by (a), and the factor $\hat m^{\,s}$ cancels in the normalized inner product. For nonzero vectors $p$ and $p'$ enclosing the angle $\varphi$,
\begin{equation*}
\|p-p'\|^2\;\ge\;2\|p\|\|p'\|(1-\cos\varphi)=\|p\|\|p'\|\,\Bigl\|\frac{p}{\|p\|}-\frac{p'}{\|p'\|}\Bigr\|^2 ,
\end{equation*}
since $\|p\|^2+\|p'\|^2\ge2\|p\|\|p'\|$. Because $\psi_s$ is invariant to positive scaling of its argument, it suffices to consider unit vectors $x$ and $x'$. With $p=A_sx$ and $p'=A_sx'$, part (d) gives $\|p-p'\|\le\|x-x'\|$ and $\|p\|\|p'\|\ge g_{\min}^{2s}$, which is the stated bound.
\end{proof}

\begin{proof}[Proof of Proposition~\ref{prop:energy}]
The mean $\bar\xi_c$ of $n$ independent $\mathcal{N}(0,\bS)$ vectors is $\mathcal{N}(0,\bS/n)$, so $\gamma=A_1\bar\xi_c\sim\mathcal{N}(0,A_1\bS A_1^{\T}/n)=\mathcal{N}(0,(m/n)I_d)$ by Proposition~\ref{prop:geometry}(a). With $a=A_1\mu_c$ we have $\|a+\gamma\|^2=\|a\|^2+2\langle a,\gamma\rangle+\|\gamma\|^2$. The cross term has mean zero and variance $4(m/n)\|a\|^2$; $\|\gamma\|^2$ is $m/n$ times a $\chi^2_d$ variable, with mean $md/n$ and variance $2m^2d/n^2$; and the two are uncorrelated because odd moments of a centered Gaussian vanish.
\end{proof}

\begin{proof}[Proof of Theorem~\ref{thm:integrity}]
The event $\|p_c+\delta-p_j\|\le\|\delta\|$ is equivalent to $D_{cj}^2+2\langle\delta,p_c-p_j\rangle\le0$, that is, to $\langle\delta,e\rangle\ge D_{cj}/2$ with the unit vector $e=(p_j-p_c)/D_{cj}$. Since $\langle\delta,e\rangle\sim\mathcal{N}(0,\lambda_c)$, its probability is $\bar\Phi(D_{cj}/(2\sqrt{\lambda_c}))=\bar\Phi(zD_{cj}/D_c)$, which is at most $\bar\Phi(z)$ because $D_{cj}\ge D_c$. The union bound over $j\ne c$ gives the first expression in \eqref{eq:union}; each of its $K-1$ terms is at most $\bar\Phi(z)$, and $(K-1)\bar\Phi(z)\le\alpha$ when $z\ge\bar\Phi^{-1}(\alpha/(K-1))$. A point at distance $a$ from the bisector crosses it when $\langle\delta,e\rangle\ge a$, with probability $\bar\Phi(a/\sqrt{\lambda_c})$. For the last statement, $A_1^{-1}=\hat m^{-1/2}\hbS^{1/2}$ by Proposition~\ref{prop:geometry}(a), so $A_1^{-1}\delta$ has covariance $\lambda_cA_1^{-1}A_1^{-\T}=(\lambda_c/\hat m)\hbS$.
\end{proof}

\begin{proof}[Proof of Theorem~\ref{thm:risk}]
Write $\lse(a)=\log\sum_j\exp(a_j)$. For example $i$ with label $y$ and strength $\lambda=\lambda_y$, put $u_j=w_j-w_y$ and $a_j=\langle u_j,\psi_i\rangle+b_j-b_y$. The noisy loss is $\lse(a+U\delta)$ with $(U\delta)_j=\langle u_j,\delta\rangle$.

(a) Concavity of the logarithm and the Gaussian moment generating function give $\E\,\lse(a+U\delta)\le\log\sum_j\exp(a_j)\,\E\exp\langle u_j,\delta\rangle=\lse(a+\kappa)$ with $\kappa_j=\tfrac\lambda2\|u_j\|^2$, which is the summand of $\mathcal{L}$. Convexity of $\lse$ and $\E\delta=0$ give $\E\,\lse(a+U\delta)\ge\lse(a)$. Since $\lse$ is nondecreasing in each argument and $\lse(a+t\mathbf{1})=\lse(a)+t$, we get $\lse(a+\kappa)-\lse(a)\le\max_j\kappa_j$. Averaging over $i$ yields the three inequalities.

(b) Each exponent $e_{ij}(W,b)=\langle w_j-w_{y_i},\psi_i\rangle+b_j-b_{y_i}+\tfrac{\lambda_{y_i}}{2}\|w_j-w_{y_i}\|^2$ is the sum of an affine function and a convex quadratic, and $\lse$ is convex and nondecreasing in each argument, so every summand of $\mathcal{L}$ is convex. The risk depends on $(W,b)$ only through the differences $w_j-w_c$ and $b_j-b_c$, and $\mathcal{L}\ge\tfrac1N\max(e_{ij},e_{i'c})$ for examples $i$ of class $c$ and $i'$ of class $j$. Here $e_{ij}+e_{i'c}$ grows quadratically in $w_j-w_c$, and $e_{ij}-e_{i'c}$ equals $2(b_j-b_c)$ up to terms in the weights alone, so $\mathcal{L}$ is coercive in the differences and attains its minimum. Let $\theta=(W,b)$ and $\theta'=(W',b')$ be minimizers and $\theta_t=(1-t)\theta+t\theta'$. Then $\mathcal{L}$ is constant on the segment, and for every $i$
\begin{align*}
\lse(e_i(\theta_t))&\le\lse\bigl((1-t)e_i(\theta)+te_i(\theta')\bigr)\\
&\le(1-t)\lse(e_i(\theta))+t\,\lse(e_i(\theta'))
\end{align*}
must hold with equality. Because $\lse$ is strictly increasing in each argument, equality on the left forces $e_{ij}(\theta_t)=(1-t)e_{ij}(\theta)+te_{ij}(\theta')$ for all $j$. The quadratic part is strictly convex in $w_j-w_{y_i}$ when $\lambda_{y_i}>0$, so $w_j-w_{y_i}=w'_j-w'_{y_i}$ for all $j$, which means $W'-W=\mathbf{1}\alpha^{\T}$ for a vector $\alpha$. Equality on the right requires $e_i(\theta')-e_i(\theta)$ to be a multiple of $\mathbf{1}$, since $\lse$ is strictly convex along every other direction. With the weight differences equal, $e_{ij}(\theta')-e_{ij}(\theta)=(b'_j-b_j)-(b'_{y_i}-b_{y_i})$, so $b'-b$ is a multiple of $\mathbf{1}$.

(c) Let $H=\{U:\sum_cu_c=0\}$ and, for $U\in H$, $G_\lambda(U)=\lambda(\mathcal{L}(U/\lambda)-\log K)$. With $a_{ij}(U)=\langle u_j-u_{y_i},\psi_i\rangle+\tfrac12\|u_j-u_{y_i}\|^2$ the exponents of $\mathcal{L}(U/\lambda)$ are $a_{ij}(U)/\lambda$. The Hessian of $\lse$ is $\operatorname{diag}(\pi)-\pi\pi^{\T}$ for a probability vector $\pi$, whose quadratic form at $a$ is a variance bounded by $\max_ja_j^2$, so $|\lse(a/\lambda)-\log K-\tfrac{1}{\lambda K}\sum_ja_j|\le\max_ja_j^2/(2\lambda^2)$. Therefore $G_\lambda\to G$ uniformly on bounded subsets of $H$, where
\begin{equation*}
G(U)=\frac{1}{NK}\sum_{i,j}a_{ij}(U)=\frac1K\sum_c\bigl(\|u_c\|^2-\langle u_c,p_c-\bar p\rangle\bigr).
\end{equation*}
The second expression uses $\sum_cu_c=0$, the equal class sizes, and the identity $\sum_{c,j}\|u_j-u_c\|^2=2K\sum_c\|u_c\|^2$ on $H$. The function $G$ is minimized on $H$ at $U^\star=\tfrac12(P-\mathbf{1}\bar p^{\T})$, which lies in $H$, and $G(U)-G(U^\star)=\tfrac1K\|U-U^\star\|_F^2$. Fix $\epsilon>0$ and let $B$ be the closed ball of radius $\epsilon$ around $U^\star$ in $H$. For $\lambda$ large enough, $\sup_B|G_\lambda-G|<\epsilon^2/(3K)$, so $G_\lambda\ge G(U^\star)+2\epsilon^2/(3K)>G_\lambda(U^\star)$ on the boundary of $B$. A convex function that is larger on the boundary of a ball than at its center attains its minimum over $H$ inside the ball. Hence $\lambda W_\lambda$, a minimizer of $G_\lambda$, satisfies $\|\lambda W_\lambda-U^\star\|_F\le\epsilon$, and since $\epsilon$ is arbitrary, $\lambda W_\lambda\to U^\star$. The classifier with weights $W_\lambda$ predicts as the one with weights $\lambda W_\lambda$, whose scores converge to $\tfrac12\langle p_c-\bar p,x\rangle$; these differ from $\tfrac12\langle p_c,x\rangle$ by a term common to all classes, so the prediction is $\argmax_c\langle p_c,x\rangle$ for large $\lambda$ whenever that maximum is attained by a single class. The limit $\lambda\to0$ is immediate from \eqref{eq:risk}.
\end{proof}

\begin{proof}[Proof of Proposition~\ref{prop:rounds}]
The carrier of $x$ decomposes as $x-m_a=\xi+(\mu_y-\mu_a)+(\mu_a-m_a)$. By Minkowski's inequality in $L^2$, $\tau_v$ is at most the sum of the root second moments of the three projections on $v$. The first is $\sigma_v$, because $\E(v^{\T}\xi)^2=\E\,\E[(v^{\T}\xi)^2\mid y]=v^{\T}\bS v$ whatever the attribution rule is. The second satisfies $\E(v^{\T}(\mu_y-\mu_a))^2=\sum_{c\ne c'}q_{cc'}(v^{\T}(\mu_c-\mu_{c'}))^2\le\varepsilon\Delta_v^2$, and the third is at most $\eta_v$. This proves \eqref{eq:leak}. If $g=1$ the separation is $\Delta_v$; otherwise $g\Delta_v=\sqrt m\,\Delta_v/\tau_v\ge\sqrt m\,\Delta_v/(\sigma_v+\eta_v+\sqrt\varepsilon\,\Delta_v)$, which is \eqref{eq:selflimit}. Under the additional assumptions, an example of class $c$ attributed to $c'$ yields the carrier $\xi+\mu_c-\mu_{c'}$ with $\xi$ independent of the attribution, so its second moment given the pair is $\bS+(\mu_c-\mu_{c'})(\mu_c-\mu_{c'})^{\T}$, and averaging with the weights $q_{cc'}$ gives $\bS+B_q$. Then $\tau_v^2=\sigma_v^2+\sum_{c\ne c'}q_{cc'}(v^{\T}(\mu_c-\mu_{c'}))^2\le\sigma_v^2+\varepsilon\Delta_v^2$, and $g\Delta_v\ge\sqrt m\,(\sigma_v^2/\Delta_v^2+\varepsilon)^{-1/2}$ when $g<1$.
\end{proof}

\bibliographystyle{elsarticle-num}
\bibliography{references}

\end{document}